\documentclass{article}

\usepackage{PRIMEarxiv}

\usepackage[utf8]{inputenc} 
\usepackage[T1]{fontenc}    
\usepackage{hyperref}       
\usepackage{url}            
\usepackage{booktabs}       
\usepackage{amsfonts}       
\usepackage{nicefrac}       
\usepackage{microtype}      
\usepackage{lipsum}
\usepackage{amsmath}
\usepackage{algorithm}%
\usepackage{algorithmicx}%
\usepackage{algpseudocode}%
\usepackage{fancyhdr}       
\usepackage{graphicx}       
\graphicspath{{media/}}     
\usepackage{rotating}
\title{GeoSSA: Geometric Sparrow Search Algorithm for UAV Path Planning and Engineering Design Optimization
\thanks{\textit{The supports provided by Macao Polytechnic University (MPU Grant no: RP/FCA-03/2022; RP/FCA-06/2022) and Macao Science and Technology Development Fund (FDCT Grant no: 0044/2023/ITP2) enabled us to conduct data collection, analysis, and interpretation, as well as cover expenses related to research materials and participant recruitment. MPU and FDCT investment in our work have significantly contributed to the quality and impact of our research findings.}: 
\textbf{Authors. Title. Pages.... DOI:000000/11111.}} 
}

\author{
  Junhao Wei \\
  Faculty of Applied Sciences \\
  Macao Polytechnic University \\
  Macao, China\\
  \texttt{p2312195@mpu.edu.mo}  
   \And
   Wenxuan Zhu \\
  Faculty of Applied Sciences \\
  Macao Polytechnic University \\
  Macao, China\\
  \texttt{p2525620@mpu.edu.mo}  
   \And
  Qingyang Xu\\
  Faculty of Applied Sciences \\
  Macao Polytechnic University \\
  Macao, China\\
  \texttt{p2311371@mpu.edu.mo}  
   \And
  Yanxiao Li\\
  Faculty of Applied Sciences \\
  Macao Polytechnic University \\
  Macao, China\\
  \texttt{P2525981@mpu.edu.mo} 
   \And
  Yifu Zhao \\
  Faculty of Applied Sciences \\
  Macao Polytechnic University \\
  Macao, China\\
  \texttt{p2523269@mpu.edu.mo}  
   \And
   Zikun Li  \\
   School of Economics and Management \\
   South China Normal University\\
   Guangzhou, China\\
   \texttt{20190731013@m.scnu.edu.cn} \\
   \And
   Ran Zhang \\
   Faculty of Applied Sciences \\
   Macao Polytechnic University \\
   Macao, China\\
  \texttt{P2512396@mpu.edu.mo}  
   \And
   Yanzhao Gu\\
   Faculty of Applied Sciences \\
   Macao Polytechnic University \\
   Macao, China\\
  \texttt{p2311998@mpu.edu.mo}  
   \And
   Jinhong Song \\
   Faculty of Applied Sciences \\
   Macao Polytechnic University \\
   Macao, China\\
  \texttt{p2315937@mpu.edu.mo}  
   \And
   Yapeng Wang \\
   Faculty of Applied Sciences \\
   Macao Polytechnic University \\
   Macao, China\\
  \texttt{yapengwang@mpu.edu.mo}  
   \And
   Zhiwen Wang \\
   School of Nursing \\
   Peking University \\
   Beijing, China\\
   \texttt{wzwjing@bjmu.edu.cn} \\
   \And
   Ngai Cheong \\
   Faculty of Applied Sciences \\
   Macao Polytechnic University \\
   Macao, China\\
   \texttt{ncheong@mpu.edu.mo} \\
   \And
   Sio-Kei Im \\
   Macao Polytechnic University \\
   Macao, China\\
   \texttt{marcusim@mpu.edu.mo} \\
   \And
   Xu Yang* \\
   Faculty of Applied Sciences \\
   Macao Polytechnic University \\
   Macao, China\\
   \texttt{xuyang@mpu.edu.mo} \\
}

\begin{document}
\maketitle

\begin{abstract}
Metaheuristic algorithms have been widely applied to complex optimization problems due to their independence from gradient information, strong global search capability, and robust performance. The Sparrow Search Algorithm (SSA), characterized by its simple structure and ease of implementation, nevertheless suffers from an insufficient balance between exploration and exploitation, making it prone to premature convergence and slow optimization progress. To address these shortcomings, this paper proposes a Geometric Sparrow Search Algorithm (GeoSSA). By integrating Good Nodes Set initialization, a Sine-Cosine Enhanced Producer position update strategy, and a Triangular-Walk Enhanced Edge Sparrow update strategy, GeoSSA significantly improves the global exploration ability, local exploitation efficiency, and convergence stability of the original SSA. To thoroughly validate the effectiveness of GeoSSA, we conducted ablation studies, qualitative analysis, and comparative experiments on 23 benchmark functions against state-of-the-art algorithms. Experimental results show that GeoSSA achieves the best or near-best performance in terms of average fitness, standard deviation, Wilcoxon tests, and Friedman rankings, with an Overall Effectiveness ($OE$) of 95.65\%. Its overall performance is significantly superior to all compared algorithms. In three-dimensional UAV path planning tasks, GeoSSA demonstrates excellent stability and superior path quality. In four categories of engineering design optimization problems, GeoSSA consistently attains the highest solution accuracy and strongest stability. GeoSSA not only exhibits outstanding global optimization performance on standard benchmark functions but also shows strong robustness and generalization ability in practical applications such as UAV path planning and engineering design. Therefore, GeoSSA provides an efficient and reliable solution framework for complex optimization problems.
\end{abstract}

\keywords{Sparrow Search Algorithm \and triangular walk \and metaheuristic \and path planning \and engineering design}

\section{Introduction}\label{sec1}
In the field of optimization, metaheuristic algorithms have received sustained attention due to their efficiency and adaptability in solving complex, nonlinear, and multimodal optimization problems. As an advanced extension of heuristic algorithms, metaheuristics typically integrate randomized mechanisms with local search strategies, leveraging the synergy between global exploration and local exploitation to search for optimal or near-optimal solutions in vast and structurally complex search spaces. In this process, exploration aims to broadly investigate the entire search domain to prevent premature stagnation in local regions, whereas exploitation focuses on intensively refining promising solution areas to accelerate convergence and enhance solution quality. Benefiting from their independence from strict mathematical models and suitability for black-box and highly complex systems, metaheuristic algorithms are capable of addressing high-dimensional, multi-constraint, and non-convex problems that are challenging for traditional optimization methods. Although these algorithms do not guarantee strict global optimality, they are typically able to produce high-quality solutions efficiently within reasonable computational time. In recent years, owing to their strong optimization capability, generalizability, and ease of implementation, metaheuristic algorithms have been widely applied across various engineering and intelligent system domains, including path planning, engineering structural design optimization, WSN coverage optimization, neural network parameter learning, and feature selection.\par
In 1995, Kennedy and Eberhart introduced Particle Swarm Optimization (PSO), inspired by the foraging behavior of birds \cite{PSO}. In 2005, Karaboga et al. proposed the Artificial Bee Colony algorithm (ABC), modeled after the foraging behavior of honeybees \cite{ABC}. In 2014, Seyedali Mirjalili et al. developed the Grey Wolf Optimizer (GWO), inspired by the leadership hierarchy and collaborative hunting mechanism of grey wolves \cite{GWO}. In 2019, Elhamifar et al. proposed Harris Hawks Optimization (HHO), based on the cooperative predation strategies of Harris hawks \cite{HHO}. In 2023, Jia et al. introduced the Crayfish Optimization Algorithm (COA), inspired by the foraging, heat dissipation, and competitive behaviors of crayfish \cite{COA}. In 2024, Fu et al. proposed the Red-billed Blue Magpie Optimizer (RBMO), inspired by the species' food-searching and predatory behaviors \cite{RBMO}. Most recently, in 2025, Wang et al. developed the Cuckoo Catfish Optimizer (CCO), modeled after the searching, predation, and brood parasitism behaviors of cuckoo catfish \cite{CCO}. \par
Despite the remarkable success of these methods over the past decades, several inherent limitations remain, including premature convergence, insufficient exploration capability, slow convergence speed, and search stagnation in high-dimensional landscapes. To address these challenges, numerous improved strategies have been proposed, such as hybrid algorithmic frameworks, dynamic balancing of exploration and exploitation, memory-based mechanisms, chaotic perturbation schemes, and more realistic nature-inspired behavioral models, thereby continuously advancing the development of metaheuristic algorithms for complex optimization tasks.\par
In 2022, Zhang et al. introduced LMRAOA, which enhances global search capability using a Multi-Leader Wandering Search strategy, improves local search efficiency via a Random High-Speed Jumping strategy, and avoids local stagnation through an adaptive lens opposition-based learning mechanism with dynamic parameter tuning \cite{LMRAOA}. In 2023, Yu et al. proposed HGWODE, a hybrid algorithm integrating GWO and Differential Evolution, which effectively balances exploration and exploitation \cite{HGWODE}. In 2024, Wei et al. incorporated tent-mapping initialization, Levy flight, and adaptive $t$-distribution into PSO, formulating IPSO, which achieves faster convergence and stronger escape capability from local optima compared with PSO \cite{IPSO}. In 2025, Lu et al. proposed MRBMO by combining Good Nodes Set initialization with Levy flight, demonstrating outstanding performance in antenna S-parameter optimization \cite{mrbmo}. Table~\ref{algs2} summarizes representative classical and emerging metaheuristic algorithms.\par

\begin{table}[htbp]
	\centering
	\caption{Research on the metaheuristic algorithms.}

	\begin{tabular}{llll}
	\toprule
	Algorithm & Year & Author & Source of inspiration \\
	\midrule
    PSO~\cite{PSO} & 1995 & Kennedy et al. & Behavior of birds.\\
    & & & \\
    ABC~\cite{ABC}  & 2005 & Karaboga et al. & Foraging behavior of honeybees   \\
     & & & \\
    GWO~\cite{GWO}  & 2014 & Seyedali Mirjalili et al. & Leadership hierarchy and collaborative   \\
     & & & hunting mechanism of grey wolves.\\
     & & & \\
    HHO~\cite{HHO}  & 2019 & Elhamifar et al. & Cooperative predation strategies of  \\
     & & &  Harris hawks.\\
     & & & \\
    COA~\cite{COA}  & 2023 & Jia et al. & Foraging, heat dissipation, and competitive  \\
     & & & behaviors of crayfish.\\
     & & & \\
    RBMO~\cite{RBMO}  & 2024 & Fu et al. & Food-searching and predatory behaviors  \\
     & & & of Red-billed Blue Magpie.\\
     & & & \\
    CCO~\cite{CCO}  & 2025 & Wang et al. &  Searching, predation, and brood parasitism  \\
     & & & behaviors of cuckoo catfish.\\
     & & & \\
    LMRAOA~\cite{LMRAOA}  & 2022 & Zhang et al. & Multi-Leader Wandering Search strategy,  \\
    & & &  LIOBL.\\
    & & & \\
    HGWODE~\cite{HGWODE}  & 2023 & Yu et al. & Grey Wolf Optimizer, Differential Evolution.\\
     & & &    \\
    IPSO~\cite{IPSO}  & 2024 & Wei et al. & Tent mapping, Levy flight, adaptive \\
     & & &  $t$-distribution  \\
	  & & &    \\
	MRBMO~\cite{mrbmo}  & 2025 & Lu et al. & Good Nodes Set, Levy flight, LIOBL. \\
	\bottomrule
	\end{tabular}
	\label{algs2}
\end{table}

The Sparrow Search Algorithm (SSA), proposed by Xue et al. in 2020, is a metaheuristic optimization method inspired by the collective intelligence of sparrow populations \cite{SSA}. SSA features a relatively simple structure, making it easy to understand and implement. However, SSA suffers from poor balance between exploration and exploitation, and the population quality tends to degrade during iterations, leading to weak global exploration and premature convergence to local optima. These limitations hinder the competitiveness of SSA in solving complex real-world optimization problems.\par
To address these issues, this paper proposes a Geometric Sparrow Search Algorithm (GeoSSA), which effectively balances exploration and exploitation through the integration of geometric initialization and enhanced update mechanisms. Experimental results demonstrate that GeoSSA successfully avoids premature convergence and achieves superior performance in UAV path planning and engineering design optimization tasks.

\section{Current Research on UAV Path Planning}
With the rapid development of the low-altitude economy, unmanned aerial vehicle (UAV) systems have witnessed growing demand across various domains, including intelligent logistics, precision agriculture, disaster inspection, and military reconnaissance. According to forecasts by the International Association for Unmanned Systems, the global commercial UAV market size is expected to exceed USD 120 billion by 2025. However, reliable path-planning capability in complex environments remains a key technological bottleneck restricting the large-scale deployment of UAVs. Existing studies indicate that approximately 23\% of potential UAV missions fail to be executed due to instability in path planning, insufficient computational efficiency, or poor adaptability to dynamic environments. In urban canyons, mountainous terrains, forested areas, and dynamically regulated airspaces, traditional path-planning approaches rely primarily on two-dimensional geometric constraint models, lacking adequate representation of three-dimensional spatial structures and time-varying obstacles. Consequently, these methods often suffer from delayed response, limited obstacle-avoidance capability, and suboptimal path quality during real-world deployment.\par
In the early development of UAV path planning, deterministic geometric planning methods dominated the research landscape, including A*, D*, Rapidly-Exploring Random Trees (RRT), and their improved variants. The A* algorithm conducts heuristic search through a cost function and has proven effective for shortest-path planning in grid-based environments. D* extends this capability by supporting environmental updates, making it suitable for partially dynamic settings. Meanwhile, RRT and its variants (e.g., RRT*) perform incremental sampling to efficiently explore high-dimensional spaces, forming an important class of algorithms for handling complex obstacle environments \cite{drrt} \cite{AHRRT}. Nevertheless, these techniques heavily rely on accurate environmental modeling and tend to experience declining search efficiency, slow path refinement, or difficulty in guaranteeing global optimality when applied to high-dimensional or strongly dynamic environments.\par
With the advancement of intelligent optimization, researchers have increasingly recognized the unique advantages of metaheuristic algorithms for UAV path planning. metaheuristic algorithms do not require explicit mathematical formulations, are well suited for complex non-convex search spaces, and naturally accommodate multi-constraint and multi-objective optimization. Thus, they can more effectively address the demands of UAV navigation in rugged terrain, unstructured environments, and scenarios involving dynamic obstacles. Extensive studies have demonstrated that metaheuristic methods such as Particle Swarm Optimization (PSO), Genetic Algorithms (GA) \cite{GA}, Ant Colony Optimization (ACO) \cite{ACO} and Grey Wolf Optimizer (GWO) can generate high-quality paths within limited computation time by balancing exploration and exploitation. These methods significantly enhance path smoothness, safety, and global optimality. Compared with traditional deterministic algorithms, metaheuristic approaches exhibit stronger robustness and flexibility in high-dimensional search, redundant path avoidance, navigation in multi-obstacle environments, and complex constraint handling, thereby becoming a prominent research direction in UAV path planning.\par
In recent years, driven by progress in deep learning and reinforcement learning, researchers have also explored learning-based frameworks for UAV path planning, such as Deep Q-Networks (DQN), imitation learning, and policy gradient methods. Although learning-based approaches reveal new opportunities concerning generalization and online decision-making in complex environments, their reliance on extensive training data, sensitivity to model stability, and limited interpretability pose significant challenges for engineering applications. For these reasons, metaheuristic algorithms continue to play a central role in UAV path planning. Particularly in scenarios requiring a balance among safety, robustness, computational efficiency, and engineering feasibility, metaheuristic approaches demonstrate irreplaceable advantages.

\section{Current Research on Engineering Design}
Before the advent of computer technology, engineering design relied predominantly on experience and manual calculations. Designers completed complex structural designs through extensive trial-and-error processes and repeated adjustments, drawing heavily on their professional knowledge and intuition. This traditional approach was not only inefficient but also susceptible to limitations imposed by the designer's personal expertise and subjective judgment, leading to uncertainty in design outcomes and an inability to guarantee optimality. With continuous advancements in science and technology, the introduction of mathematical methods marked a new stage in engineering design. Designers began to employ mathematical models and equations to describe design problems more precisely-such as using finite element analysis (FEA) to evaluate structural strength and stability, or applying optimization techniques to search for optimal design solutions. Nonetheless, these approaches still relied on manual derivations and computations, which were time-consuming, resource-intensive, and restricted to relatively simple design problems.\par
The rapid development of computer technology ushered engineering design into the digital era. The integration of computers significantly accelerated the design process and enabled complex calculations and simulations to become feasible. Engineers began utilizing computer-aided design (CAD), computer-aided engineering (CAE), and computer-aided manufacturing (CAM) systems to support design, analysis, and production workflows. Computers not only facilitated large-scale numerical computations and simulations but also empowered designers to explore broader design spaces. However, as engineering problems grew increasingly complex with more intricate constraints, traditional computational methods faced bottlenecks in computational resources and optimization efficiency. For multi-objective, nonlinear, or highly constrained engineering design problems, computer-aided design tools alone remained insufficient to consistently deliver optimal solutions.\par
In the 21st century, the rise of metaheuristic algorithms brought another major breakthrough to engineering design optimization. Metaheuristic algorithms—such as Genetic Algorithms (GA), Particle Swarm Optimization (PSO), Ant Colony Optimization (ACO), and Simulated Annealing (SA) \cite{SA}—draw inspiration from biological swarm behaviors or physical processes, enabling efficient searches within complex design spaces to identify global or near-global optima. Unlike traditional mathematical optimization techniques, metaheuristic algorithms do not rely on explicit analytical formulations and can handle high-dimensional, nonlinear, multi-objective, and constrained design problems while avoiding premature convergence to local minima. As a result, they have been widely applied to structural optimization, mechanical design, electronic circuit design, aerospace engineering, and many other domains, becoming essential tools for solving complex engineering design tasks. Their strong global search capability and flexibility are particularly advantageous in large-scale, multi-variable, and difficult-to-model design scenarios.\par
With the rapid progress of machine learning and artificial intelligence, an increasing number of engineering design problems have begun adopting learning-based optimization approaches. Techniques such as deep learning and reinforcement learning enable computational systems to autonomously learn from large volumes of data and experience, thereby improving their ability to address highly complex design challenges. Machine learning methods can perform design optimization through data modeling and pattern recognition and can dynamically adjust optimization strategies to achieve adaptive design decision-making. However, despite their promising capabilities, machine learning approaches face challenges including heavy dependence on large datasets, high model complexity, and expensive training costs. Consequently, while machine learning provides unique advantages for certain design tasks, it cannot fully replace traditional optimization techniques, especially in scenarios involving limited samples, high-dimensional search spaces, or complex constraints.\par
Against this background, metaheuristic algorithms continue to play a crucial role in engineering design optimization. Compared to machine learning, metaheuristic algorithms do not require large training datasets and can effectively handle uncertainty and various types of constraints. When addressing complex multi-objective optimization tasks or navigating high-dimensional design spaces, metaheuristic algorithms balance global exploration and local exploitation, thereby avoiding the risks of local stagnation and overfitting while offering strong robustness and flexibility. These advantages are particularly significant in the early stages of design, where sample data are often scarce. Therefore, although machine learning methods hold valuable potential in certain application fields, metaheuristic algorithms remain indispensable tools for engineering design optimization—especially when tackling complex, highly constrained, or dynamically changing optimization problems, where their distinctive strengths become even more evident.

\section{SSA}
The Sparrow Search Algorithm (SSA), which was proposed by Xue et al. in 2020, is an innovative population-based optimization method inspired by the foraging and anti-predation behaviors of sparrow flocks \cite{SSA}. By simulating the natural behaviors of sparrows, SSA is capable of performing efficient global search and finding optimal solutions. SSA introduces two main roles: producers and scroungers, while also adopting a feedback-based strategy to optimize the balance between exploration and exploitation.\par
The original SSA algorithm initializes the population using a pseudo-random number method:\par
\begin{equation}
    {X}_{i,j}=(ub-lb)\cdot Rand+lb
    \label{eq1}
\end{equation}
where ${X}_{i,j}$ denotes the position of the $i^{th}$ sparrow in the $j^{th}$ dimension; $ub$ and $lb$ represent the upper and lower bounds of the search space; $Rand$ is a random number uniformly distributed within [0,1].\par
In SSA, producers (the current optimal solution), which have higher fitness values, are more likely to lead the flock in searching for food. Since producers are responsible for foraging and guiding the entire population's movement, they are capable of searching for food in a broader area compared to scroungers. In each iteration, the position of the producer is updated as follows:\par
\begin{equation}
    X_{t+1}^{i, j}=\left\{\begin{array}{ll}
X_{t}^{i, j} \cdot \exp \left(-i \cdot \frac{\alpha}{T}\right) & \text { if } R_{2}<S T \\
X_{t}^{i, j}+Q \cdot L & \text { if } R_{2} \geq S T
\end{array}\right.
    \label{eq2}
\end{equation}
where $X_{t}^{i, j}$ represents the current position of the sparrow in the $j^{th}$ dimension at iteration $t$; $T$ is the maximum number of iterations; $a \in(0,1]$ is a random number used to control the range of the search; $R_2 \in[0,1]$ is the alarm value; $ST \in[0.5,1]$ is the safety threshold; $Q$ is a random number following a normal distribution; $L$ is a matrix with all elements equal to 1.\par
When $R_2<ST$, it indicates that no predator has been detected, and producers enter a broad-range search mode. Conversely, when $R_2\geq ST$, predators are perceived, and the producers—along with other members of the population—rapidly relocate to a safer region.\par
Scroungers acquire food by monitoring the position of the producers. If a scrounger successfully finds a better food source, it will update its position according to the following rule:\par
\begin{equation}
    X_{t+1}^{i,j}=
    \begin{cases}
        Q \cdot \exp\left(\frac{X_t^{worst} - X_t^{i,j}}{i^2}\right) & \text{if } i > \frac{n}{2} \\
        X_{t+1}^P + |X_t^{i,j} - X_{t+1}^P| \cdot A^+ \cdot L & \text{otherwise}
    \end{cases}
    \label{eq3}
\end{equation}
where $X^{worst}$ represents the current global worst position; $X^P$ is the best position of the producer; $A$is a matrix with elements either 1 or -1, randomly assigned to control the direction; $L$ is a matrix consisting of ones, used to adjust the step size of the forager.\par
If $i>\frac{n}{2}$, it indicates that the current scrounger has relatively poor fitness and may be in a starving state. Consequently, it needs to search for new food sources more frequently.
When sparrows are located at the edge of the flock, they will fly towards a safe area to avoid predator attacks. The update rule for this behavior is as follows:\par
\begin{equation}
    X_{t+1}^{i,j} =
    \begin{cases}
        X_{{best}} + \beta \cdot |X_t^{i,j} - X_{{best}}| & \text{if } f_i > f_g \\
        X_t^{i,j} + K \cdot \left( |X_t^{i,j} - X_t^{{worst}}| \cdot \frac{(f_i - f_w)}{f_i} + \epsilon \right) & \text{if } f_i = f_g
    \end{cases}
    \label{eq4}
\end{equation}
where $X_{best}$ is the current global best position; $f_i$ is the fitness value of the current sparrow, and $f_g$ is the global best fitness; $\beta$ is a random distribution constant that controls the step size; $K \in [-1,1]$ is a random number that determines the direction of the step; $\epsilon$ is a constant used to prevent division by zero errors.\par
The behavioral strategy of edge sparrows is determined by comparing their fitness with the global best fitness to adjust their positions. These individuals tend to move closer to the center of the group to increase their safety.\par
The pseudocode of the original SSA is presented in Algorithm~\ref{alg1}.
\begin{algorithm}
    \caption{Sparrow Search Algorithm} 
    \begin{algorithmic}
        \State \textbf{Begin}
		\State Input:
		\State $T$: the maximum iterations;
		\State $PD$: the number of producers;
		\State $SD$: the number of sparrows who perceive the danger;
		\State $R_2$: tthe alarm value;
		\State $n$: the number of sparrows;
        \State Initialize a population of $n$ sparows and define its relevant parameters.
        \State \hspace{1em} \textbf{While} $t < T$
            \State \hspace{2em} Rank the fitness values and find the current best individual and the current worst individual;
				\State \hspace{3em} \textbf{for} $i=1:PD$
				\State \hspace{4em} Using Eq.~\ref{eq2} update the sparrow's location;
				\State \hspace{3em} \textbf{end for} 
				
                \State \hspace{3em} \textbf{for} $i=(PD+1):n$
				\State \hspace{4em} Using Eq.~\ref{eq3} update the sparrow's location;
				\State \hspace{3em} \textbf{end for}

				\State \hspace{3em} \textbf{for} $l=1:SD$
				\State \hspace{4em} Using Eq.~\ref{eq4} update the sparrow's location;
				\State \hspace{3em} \textbf{end for}

            \State \hspace{2em} Get the current new location;
            \State \hspace{2em} If the new location is better than before,update it;            
            \State \hspace{2em} $t = t + 1$
        \State \hspace{1em} \textbf{end while}
        \State \textbf{return} $X_{best}$, $f_g$
        \State \textbf{End}
    \end{algorithmic}
	\label{alg1} 
\end{algorithm}

\section{The Proposed GeoSSA}
\subsection{Good Nodes Set Initialization}
\begin{figure}[t!]
    \centering
    \includegraphics[width=\textwidth]{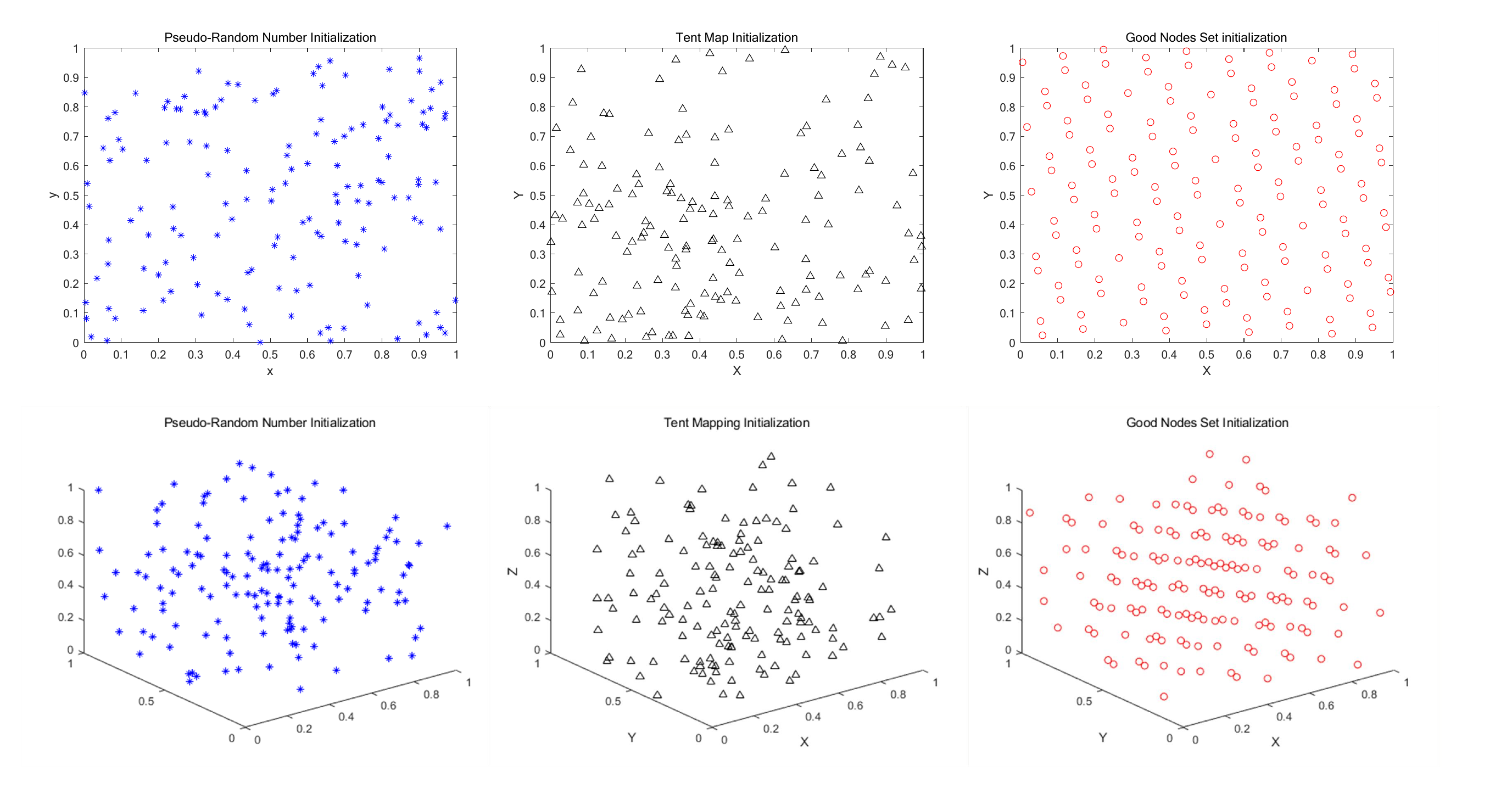}
    \caption{Comparison of different initialization methods. (a) Population initialized using Pseudo-Random Numbers when $N$=150; (b) Population initialized using Tent Mapping when $N$=150; (c) Population initialized using Good Nodes Set method when $N$=150.} 
    \label{ini}
\end{figure}

The traditional Sparrow Search Algorithm (SSA) employs Pseudo-Random Numbers initialization to generate the population. Although Pseudo-Random Numbers initialization is straightforward and exhibits strong randomness, the resulting individuals are often unevenly distributed across the search space. This uneven distribution can lead to population clustering effects, thereby reducing population diversity. Consequently, SSA is prone to premature convergence and diminished exploration efficiency during the search process. As illustrated in the left panel of Fig.~\ref{ini}, when the population size is set to $N$=150, the Pseudo-Random Numbers initialization produces a distribution with clearly observable clustered regions and large vacant areas.\par
To address the limitations of Pseudo-Random Numbers initialization, researchers have proposed various chaotic-mapping-based strategies, such as Tent mapping shown in in the middle panel of Fig.~\ref{ini}. However, Tent mapping is essentially a Pseudo-Random Numbers distribution approach derived from chaotic sequences, and its uniformity cannot be strictly guaranteed, particularly in high-dimensional spaces.\par
To overcome the above limitations, GeoSSA adopts the Good Nodes Set method for population initialization \cite{glnwoa} \cite{mrbmo} \cite{nawoa}. The concept of the Good Nodes Set was introduced by Hua Luogeng, aiming to generate point sets with more uniform spatial distribution. A key advantage of this method is that its construction is dimension-independent—meaning that it maintains uniformity not only in two-dimensional spaces but also in high-dimensional environments. As shown on the right side of Fig.~\ref{ini}, the population generated by the Good Nodes Set exhibits a more homogeneous distribution compared to Pseudo-Random Number initialization. This effectively avoids population clustering, increases coverage of the search space, and strengthens global exploration performance during early iterations.\par
Let $U^D$ denote the unit hypercube in a $D$-dimensional Euclidean space. The Good Nodes Set can be defined as:
\begin{equation}
    P_r^M=\{p(k)=(\{kr\},\{kr^2\},...,\{kr^D\})|k=1,2,...,M\}
    \label{eq5}
\end{equation}
where $\{x\}$ denotes the fractional part of $x$, $M$ is the number of nodes, $r>0$ is an offset parameter, $C(r,\varepsilon)$ is a constant depending on $r$, and $\varepsilon>0$ is a given constant.\par
When mapping the Good Nodes Set to the actual search space, suppose the lower and upper bounds of the $i^{th}$ dimension are $x_i^{min}$ and $x_i^{max}$, respectively. The mapping is performed as follows:
\begin{equation}
   x^i_{k}=x^i_{min}+p_i(k) \cdot (x^i_{max}-x^i_{min})
    \label{eq6}
\end{equation}

\subsection{Sine-Cosine Enhanced Producer Position Update}
In the Sparrow Search Algorithm (SSA), the traditional position-update mechanism for producer individuals relies on simple exponential decay functions or Gaussian-based updates. However, this strategy depends too heavily on local search, causing individuals to easily fall into local optima. This issue becomes more pronounced in high-dimensional or complex search spaces, where the global exploration capability of the algorithm is significantly limited.\par
To overcome this limitation, this study proposes a new producer position update strategy, called Sine-Cosine Enhanced Producer Position Update strategy \cite{gwoa}. By integrating both sine and cosine functions, this strategy enhances the position-update mechanism of producers and improves the global search ability of the traditional SSA. Specifically, the Sine-Cosine Strategy introduces a decision logic based on the value of $r_2$: when $r_2$ is smaller than the predefined threshold $ST$, the sine function is used to control the producer's position update; when $r_2$ is greater than $ST$, the cosine function is applied instead.\par
The core formula of this strategy is as follows:\par
\begin{equation}
   X^{i,j}_{t+1} = \begin{cases}
   \omega \cdot X^{i,j}_t + r_1 \cdot \sin(r_2) \cdot |r_3 \cdot X_{{best}} - X^{i,j}_t|, & \text{if } R_2 < ST \\
   \omega \cdot X^{i,j}_t + r_1 \cdot \cos(r_2) \cdot |r_3 \cdot X_{{best}} - X^{i,j}_t|, & \text{if } R_2 \geq ST
   \end{cases}
   \label{eq7}
\end{equation}
where $\omega$ is the inertia weight based on the sigmoid function; $r_1$ and $r_2$ are random numbers; $X_{best}$ is the current global best position; $r_3$ is a coefficient that adjusts the global exploration range; and $X_{{best}}$ represents the current global best position. \par
The update rule for the inertia weight $\omega$ is as follows:\par
\begin{equation}
    \omega=\frac1{1+e^{-25(\frac tT-0.5)}}
    \label{eq8}
\end{equation}
where $t$ is the current iteration count; $T$ is the maximum number of iterations.

\subsection{Triangular Walk Enhanced Edge Sparrow Position Update}
In the traditional SSA, the position update of edge sparrows (i.e., local optima individuals) typically relies on simple shifts based on the global optimal solution and random disturbances. While this position update mechanism can guide edge sparrows toward the global optimum, its update process is overly simplistic, making it prone to getting trapped in local optima. Furthermore, it lacks sufficient diversity and randomness in complex problems, leading to slow convergence and premature convergence of the algorithm.\par
\begin{figure}[htbp]
    \centering
    \includegraphics[width=0.6\textwidth]{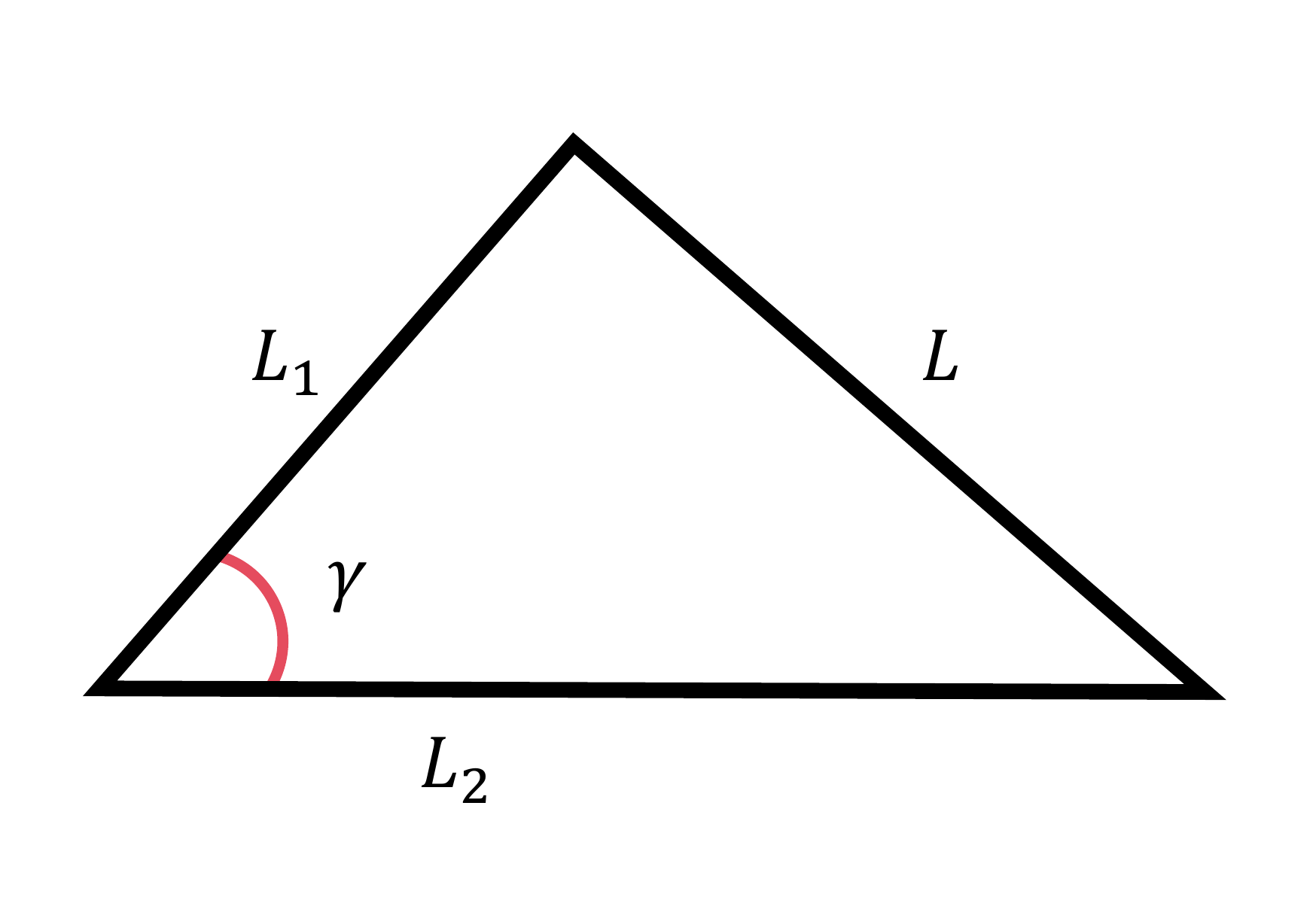}
    \caption{Adjustment factor $\alpha$ of Triangular Walk.} 
    \label{tri}
\end{figure}

To address this drawback, we proposed Triangular Walk Enhanced Edge Sparrow Position Update strategy. This strategy introduces the Triangular Walk to improve the position update of edge sparrows, enhancing the local search capability and diversity \cite{tswoa}. Specifically, the Triangular Walk is combined with the original position update mechanism to adjust the update strategy. The fundamental idea is to introduce a dynamic adjustment factor $rg$ during the update process, which is defined as:\par
\begin{equation}
   rg=0.1-0.1 \cdot \frac{t}{T}
    \label{eq11}
\end{equation}
\begin{equation}
   r=rg \cdot Rand
    \label{eq11.5}
\end{equation}
Then, we introduce a random disturbance mechanism based on the Triangular Walk during the position update, as follows:\par
\begin{equation}
   L=X_{best}- X^{i,j}_{t}
    \label{eq12}
\end{equation}
\begin{equation}
   LP=L \cdot Rand
    \label{eq13}
\end{equation}
\begin{equation}
   \alpha=L^2+LP^2-2\cdot L\cdot LP\cdot\cos\left(2\pi\cdot Rand\right)
    \label{eq14}
\end{equation}
\begin{equation}
   X^{i,j}_{t+1}=X_{t+1}^P \cdot L + r \cdot \alpha
    \label{eq15}
\end{equation}
where $L$ represents the difference between the local optimal solution and the current individual; $LP$ is a random disturbance; $\alpha$ is a new position adjustment factor calculated using trigonometric functions, shown in Fig.~\ref{tri}; and $r$ is a random coefficient controlled by $rg$.\par
Compared with traditional position update strategies, the Triangular Walk Enhanced Edge Sparrow Position Update mechanism offers significant advantages. First, the introduction of triangular walk increases the randomness of the position update process, breaking the limitation of single-direction perturbations. This allows the algorithm to move more flexibly across the solution space and reduces the likelihood of becoming trapped in local optima due to overly localized updates. Second, the dynamically adjusted factor $rg$ ensures wide-range exploration during the early stages of iteration, while gradually guiding the algorithm toward convergence in later stages, thereby improving convergence speed. Finally, the triangular-walk-based update mechanism substantially enhances the global search capability. In particular, when dealing with high-dimensional and complex optimization problems, it demonstrates stronger robustness and superior optimization performance compared with the traditional SSA.\par
The Triangular Walk Enhanced Edge Sparrow Position Update strategy strengthens the position update process of locally optimal individuals by incorporating the triangular-walk mechanism, enabling the sparrow search algorithm to maintain a better balance between global exploration and local exploitation. This significantly improves solution quality and convergence efficiency, and its advantages become especially prominent in complex, high-dimensional optimization tasks.\par
The pseudocode of GeoSSA is presented in Algorithm~\ref{alg2}.
\begin{algorithm}
    \caption{The proposed GeoSSA} 
    \begin{algorithmic}
        \State \textbf{Begin}
		\State Input:
		\State $T$: the maximum iterations;
		\State $PD$: the number of producers;
		\State $SD$: the number of sparrows who perceive the danger;
		\State $R_2$: tthe alarm value;
		\State $n$: the number of sparrows;
        \State Initialize a population of $n$ sparows using Good Nodes Set method and define its relevant parameters.
        \State \hspace{1em} \textbf{While} $t < T$
            \State \hspace{2em} Rank the fitness values and find the current best individual and the current worst individual;
				\State \hspace{3em} \textbf{for} $i=1:PD$
				\State \hspace{4em} Using Eq.~\ref{eq7} update the sparrow's location;
				\State \hspace{3em} \textbf{end for} 
				
                \State \hspace{3em} \textbf{for} $i=(PD+1):n$
				\State \hspace{4em} Using Eq.~\ref{eq3} update the sparrow's location;
				\State \hspace{3em} \textbf{end for}

				\State \hspace{3em} \textbf{for} $l=1:SD$
				\State \hspace{4em} Using Eq.~\ref{eq15} update the sparrow's location;
				\State \hspace{3em} \textbf{end for}

            \State \hspace{2em} Get the current new location;
            \State \hspace{2em} If the new location is better than before,update it;            
            \State \hspace{2em} $t = t + 1$
        \State \hspace{1em} \textbf{end while}
        \State \textbf{return} $X_{best}$, $f_g$
        \State \textbf{End}
    \end{algorithmic}
	\label{alg2} 
\end{algorithm}

\section{Experiments}
The experiments in this study were conducted on a system equipped with Windows 11 (64-bit), an Intel(R) Core(TM) i5-8300H CPU @ 2.30 GHz processor, 8 GB RAM, and MATLAB R2023a as the simulation platform. To evaluate the performance and effectiveness of the proposed GeoSSA , the following experiments were designed:\par
\begin{itemize}
	\item Experiment 1: An ablation study in which the proposed improvement strategies were removed from GeoSSA and evaluated independently on the 23 benchmark functions shown in Table~\ref{table1};
	\item Experiment 2: A qualitative analysis experiment in which GeoSSA was applied to 23 benchmark functions to comprehensively assess its performance, robustness, and exploration-exploitation balance. The evaluation includes analysis of convergence behavior, population diversity, and exploration-exploitation dynamics;
	\item Experiment 3: A comparative experiment between GeoSSA and several standard metaheuristic algorithms, advanced SSA variant, and state-of-the-art (SOTA) improved metaheuristic algorithms on the 23 benchmark functions.
\end{itemize}

\begin{table}[htbp]
    \centering
    \caption{Standard Benchmark Functions~\cite{CEC}.}
    \begin{tabular}{c c c c c}
    \hline
    Function & Function's Name & Type & Dimension & Best Value \\
    \hline
    F1 & Sphere & Uni-modal, Scalable & 30 & 0 \\
    F2 & Schwefel's Problem 2.22 & Uni-modal, Scalable & 30 & 0 \\
    F3 & Schwefel's Problem 1.2 & Uni-modal, Scalable  & 30 & 0 \\
    F4 & Schwefel's Problem 2.21 & Uni-modal, Scalable  & 30 & 0 \\
    F5 & Generalized Rosenbrock's Function & Uni-modal, Scalable  & 30 & 0 \\
    F6 & Step Function & Uni-modal, Scalable  & 30 & 0 \\
    F7 & Quartic Function & Uni-modal, Scalable  & 30 & 0 \\
    F8 & Generalized Schwefel's Function & Multi-modal, Scalable & 30 & -12569.5 \\
    F9 & Generalized Rastrigin's Function & Multi-modal, Scalable & 30 & 0 \\
    F10 & Ackley's Function & Multi-modal, Scalable & 30 & 0 \\
    F11 & Generalized Griewank's Function & Multi-modal, Scalable & 30 & 0 \\
    F12 & Generalized Penalized Function 1 & Multi-modal, Scalable & 30 & 0 \\
    F13 & Generalized Penalized Function 2 & Multi-modal, Scalable & 30 & 0 \\
    F14 & Shekel's Foxholes Function & Multi-modal, Unscalable & 2 & 0.998 \\
    F15 & Kowalik's Function & Composition, Unscalable & 4 & 0.0003075 \\
    F16 & Six-Hump Camel-Back Function & Composition, Unscalable  & 2 & -1.0316 \\
    F17 & Branin Function & Composition, Unscalable & 2 & 0.398 \\
    F18 & Goldstein-Price Function & Composition, Unscalable & 2 & 3 \\
    F19 & Hartman's Function 1 & Composition, Unscalable & 3 & -3.8628 \\
    F20 & Hartman's Function 2 & Composition, Unscalable & 6 & -3.32 \\
    F21 & Shekel's Function 1 & Composition, Unscalable & 4 & -10.1532 \\
    F22 & Shekel's Function 2 & Composition, Unscalable & 4 & -10.4029 \\
    F23 & Shekel's Function 3 & Composition, Unscalable & 4 & -10.5364 \\
    \hline
    \label{table1}
    \end{tabular}
\end{table}

\subsection{Ablation study}
In the ablation study, the three proposed strategies were individually removed or replaced with their original SSA counterparts:\par
\begin{itemize}
	\item GeoSSA1: GeoSSA1 denotes the variant in which the Good Nodes Set Initialization is replaced with pseudo-random initialization.
	\item GeoSSA2: GeoSSA2 denotes the variant in which the proposed Sine-Cosine Enhanced Producer Position Update is replaced with the original producer update strategy.
	\item GeoSSA3: GeoSSA3 denotes the variant in which the Triangular Walk Enhanced Edge Sparrow Position Update is replaced with the original edge-sparrow update mechanism.
\end{itemize}

\begin{figure}[htbp]
    \centering
    \includegraphics[width=\textwidth]{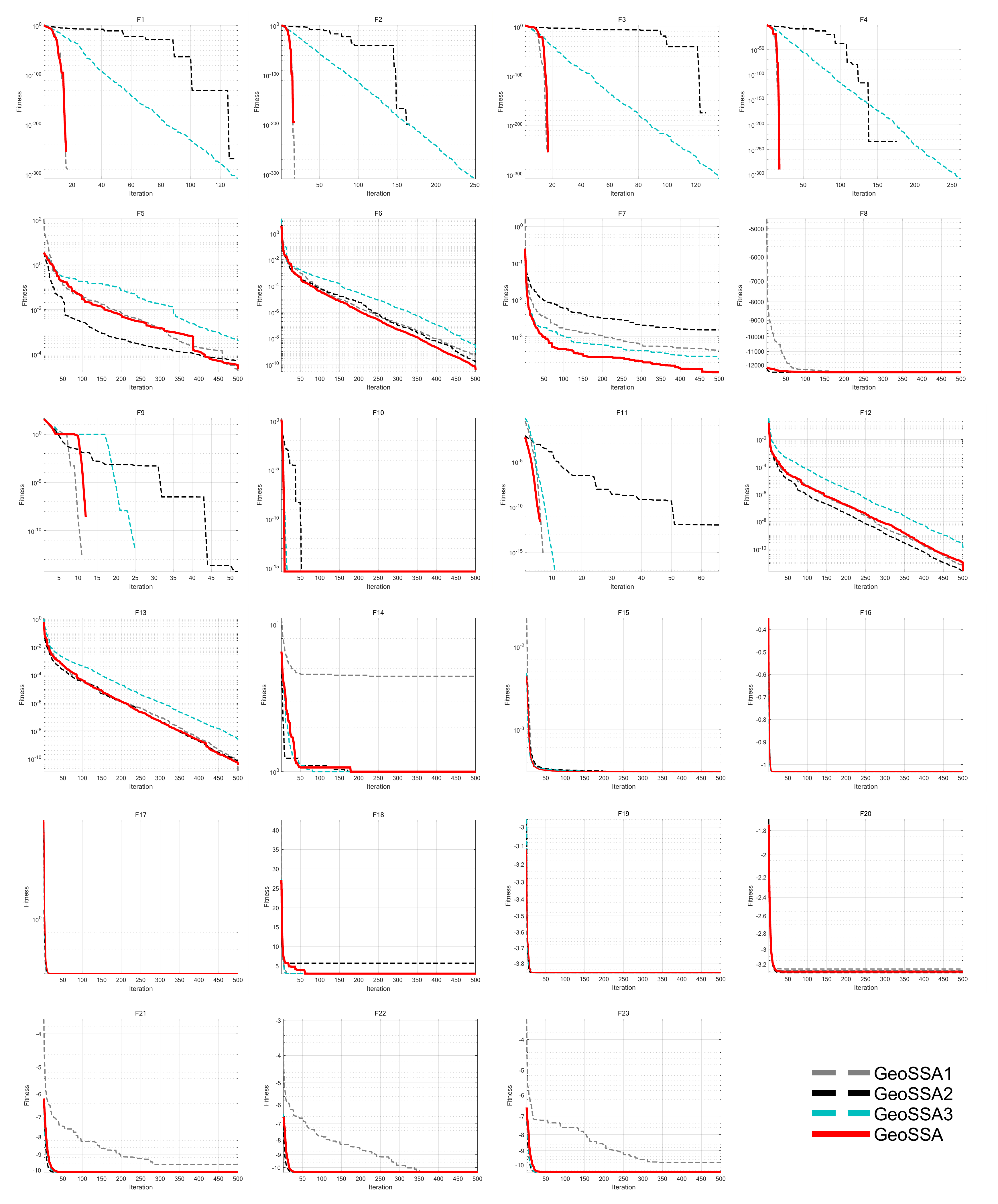}
    \caption{Iteration curves of the GeoSSAs in ablation study}
    \label{abla1}
\end{figure}

For consistency, the number of iterations was set to $T$=500 and the population size to $N$=30. Each algorithm was independently executed 30 times on the 23 benchmark functions to validate the effectiveness of each proposed strategy. The experimental results are presented in Fig.~\ref{abla1} and Table 3.\par
The Friedman test results demonstrate that GeoSSA ranks first among all variants in the ablation study. Moreover, the results clearly show that each of the proposed modules contributes significantly to improving the performance of GeoSSA.\par

\begin{table}[htbp]
\centering
\caption{Friedman Test Results of the Algorithms in Ablation Study. $AFV$ indicates Average Friedman Value.}
\begin{tabular}{ccccc}
\toprule
Functions & GeoSSA1 & GeoSSA2 & GeoSSA3 & GeoSSA \\
\midrule
F1  & 2.5000 & 2.5000 & 2.5000 & 2.5000 \\
F2  & 2.5000 & 2.5000 & 2.5000 & 2.5000 \\
F3  & 2.5000 & 2.5000 & 2.5000 & 2.5000 \\
F4  & 2.5000 & 2.5000 & 2.5000 & 2.5000 \\
F5  & 2.5667 & 2.0333 & 3.4667 & 1.9333 \\
F6  & 2.2667 & 2.1000 & 3.6000 & 2.0333 \\
F7  & 2.6333 & 3.4333 & 1.6333 & 2.3000 \\
F8  & 3.4000 & 2.6000 & 2.7333 & 1.2667 \\
F9  & 2.5000 & 2.5000 & 2.5000 & 2.5000 \\
F10 & 2.5000 & 2.5000 & 2.5000 & 2.5000 \\
F11 & 2.5000 & 2.5000 & 2.5000 & 2.5000 \\
F12 & 2.0667 & 2.2667 & 3.3333 & 2.3333 \\
F13 & 2.1000 & 2.2000 & 3.5333 & 2.1667 \\
F14 & 2.7167 & 2.2167 & 2.9000 & 2.1667 \\
F15 & 2.4333 & 2.6333 & 2.5667 & 2.3667 \\
F16 & 2.4667 & 2.3167 & 2.7667 & 2.4500 \\
F17 & 2.5000 & 2.5000 & 2.5000 & 2.5000 \\
F18 & 2.2167 & 2.8333 & 2.3667 & 2.5833 \\
F19 & 2.4500 & 2.3333 & 2.8167 & 2.4000 \\
F20 & 2.9000 & 2.1333 & 2.7000 & 2.2667 \\
F21 & 2.9167 & 2.0333 & 2.3167 & 2.7333 \\
F22 & 2.9000 & 2.0500 & 2.2833 & 2.7667 \\
F23 & 2.8833 & 2.1833 & 2.3833 & 2.5500 \\\midrule
$AFV$ & 2.5616 & 2.4072 & 2.6696 & \textbf{2.3616} \\\midrule
Rank & 3 & 2 & 4 & \textbf{1} \\
\bottomrule
\end{tabular}
\end{table}

First, the Good Nodes Set Initialization generates a more uniformly distributed sparrow population in the initial stage, expanding the coverage of the search space and preventing the formation of population clusters that would otherwise create search blind spots. This mechanism exhibits strong global exploration capability, particularly for complex multimodal functions (e.g., F20-F23), enabling the algorithm to more effectively identify promising regions. Second, the Sine-Cosine Enhanced Producer Position Update substantially improves search diversity during the exploration phase, allowing producers to make wider jumps in the solution space and rapidly locate the global optimum. This results in faster convergence on functions such as F1-F4 and F9-F11. The incorporation of sine and cosine operators enhances the algorithm's exploratory behavior, enabling effective avoidance of persistent local optima. Furthermore, this update mechanism considers not only the distance between individuals and the global optimum but also the relative distances among individuals, thereby further reinforcing global search capability. Thus, GeoSSA demonstrates superior performance on complex functions such as F5-F7. Finally, the Triangular Walk Enhanced Edge Sparrow Position Update introduces additional randomness while its dynamic factor ensures a balanced transition between global exploration and local exploitation. This allows the algorithm to converge quickly on simple functions (e.g., F1-F4) and effectively escape local optima on more difficult functions (e.g., F5-F7 and F12-F13). The three improvement modules complement one another, jointly enhancing global exploration, local exploitation, and population diversity maintenance. As a result, GeoSSA exhibits faster convergence and higher solution accuracy across different categories of benchmark functions, fully validating the effectiveness and robustness of the proposed strategies.

\subsection{Qualitative Analysis}
\begin{figure}[htbp]
    \centering
    \includegraphics[width=0.9\textwidth]{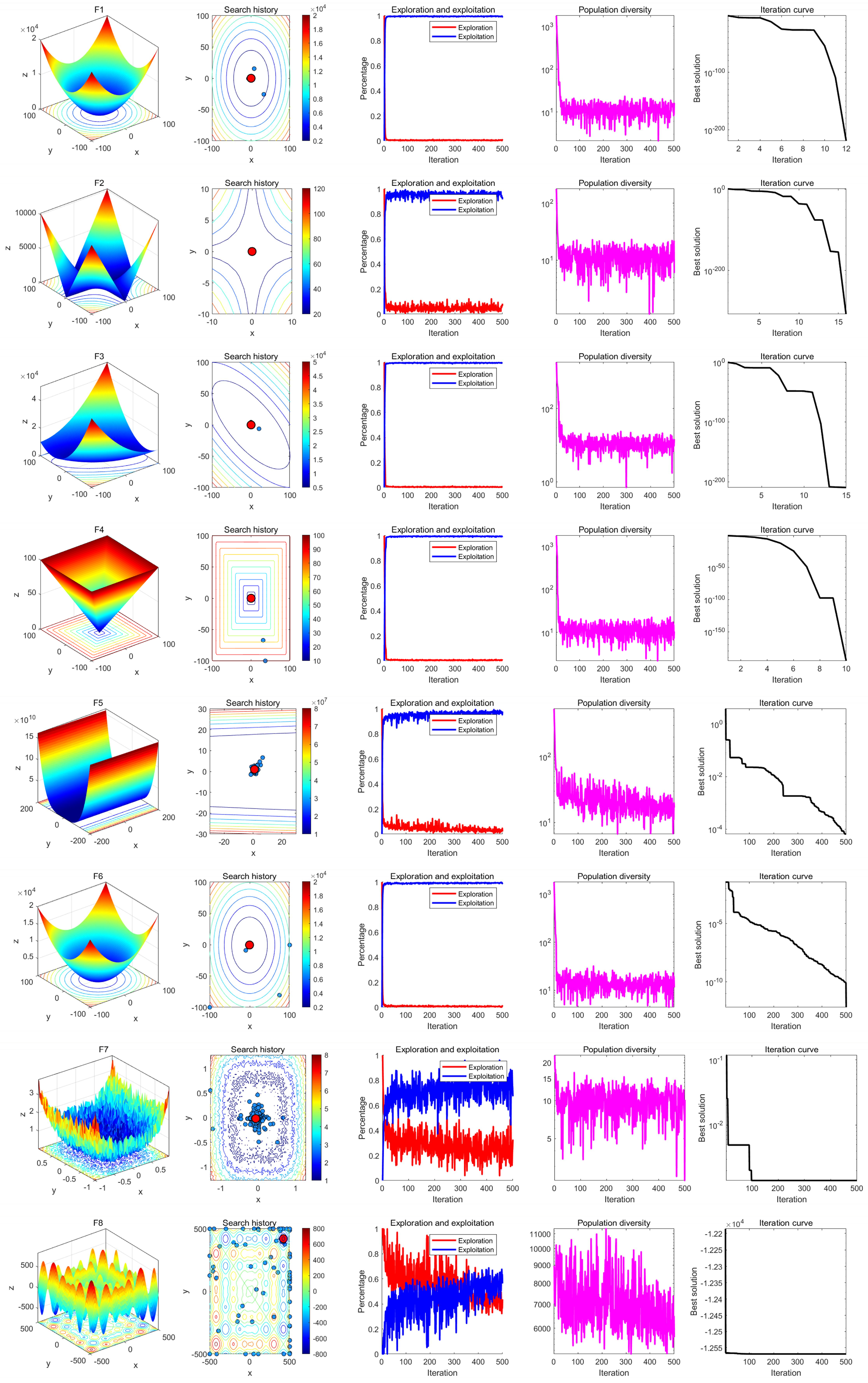}
    \caption{Results of qualitative analysis experiment (F1-F8).}
    \label{1-8}
\end{figure}

\begin{figure}[htbp]
    \centering
    \includegraphics[width=0.9\textwidth]{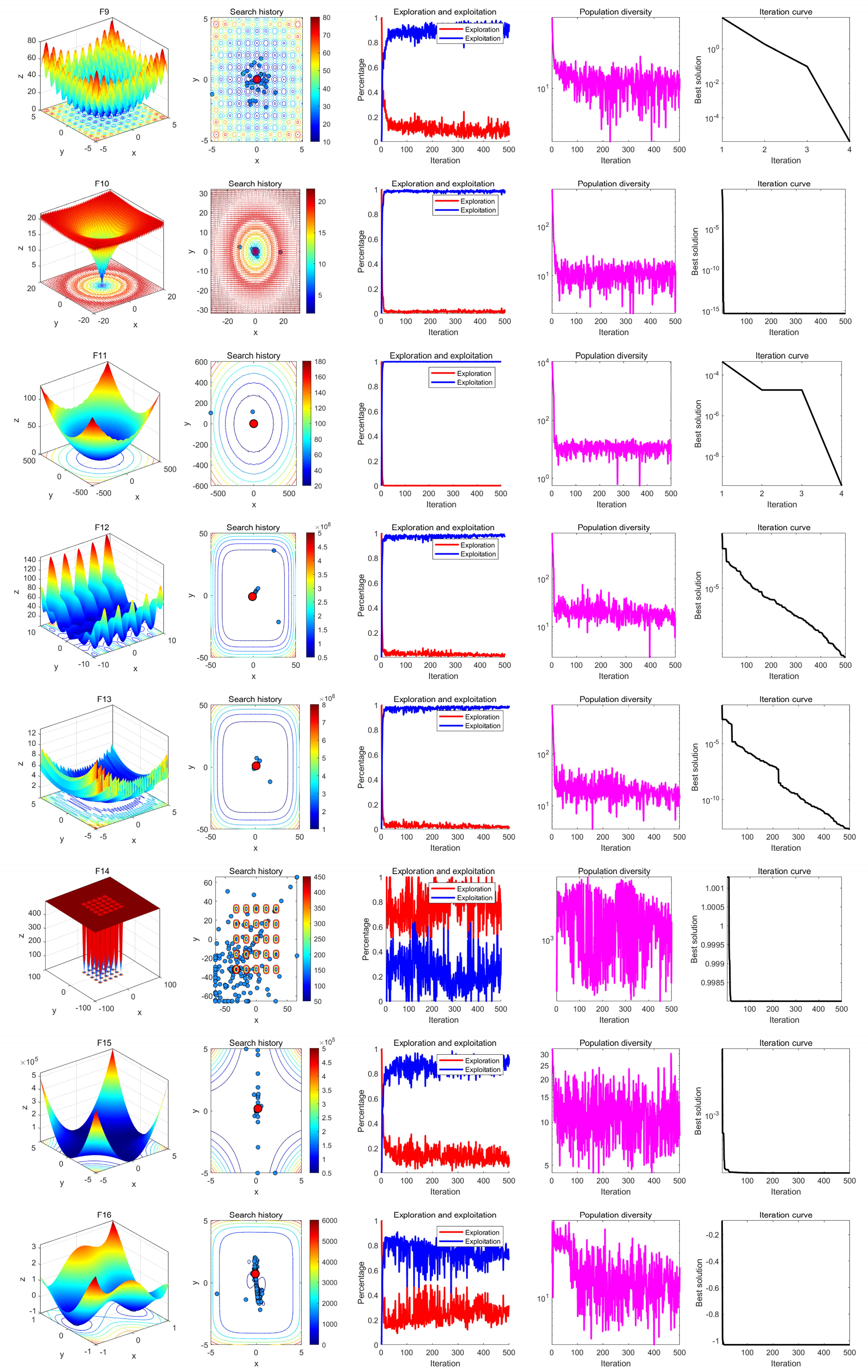}
    \caption{Results of qualitative analysis experiment (F9-F16).}
    \label{9-16}
\end{figure}

\begin{figure}[htbp]
    \centering
    \includegraphics[width=0.9\textwidth]{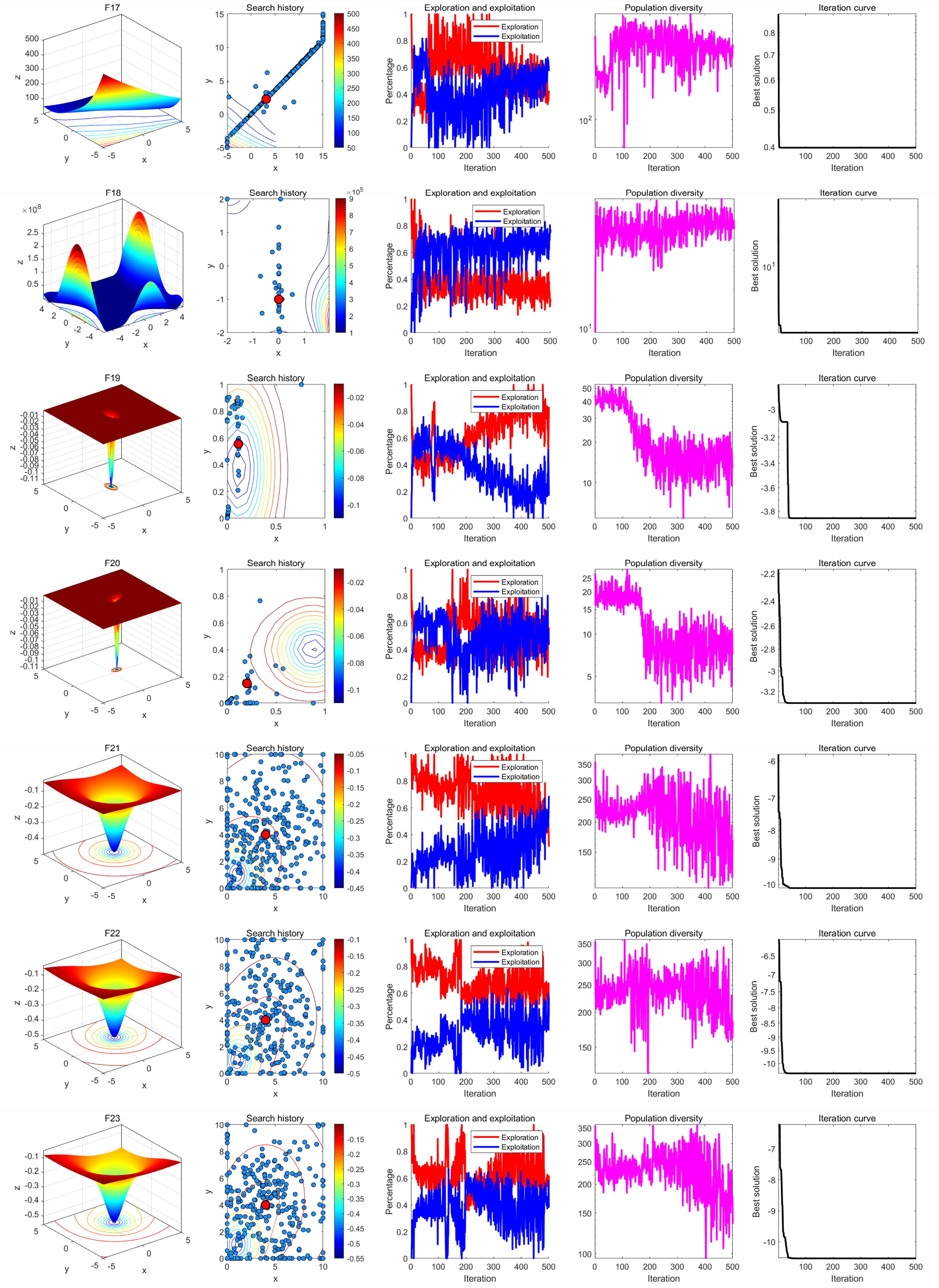}
    \caption{Results of qualitative analysis experiment (F17-F23).}
    \label{17-23}
\end{figure}

In the qualitative analysis experiment, the number of iterations was set to $T$=500 and the population size to $N$=30. GeoSSA was independently executed on the 23 benchmark functions listed in Table~\ref{table1} to analyze its search history, exploration-exploitation ratio, and population diversity. To facilitate comparison and interpretation, function landscapes and convergence curves were also provided. The qualitative analysis results are shown in Fig.~\ref{1-8}, Fig.~\ref{9-16}, Fig.~\ref{17-23} and include the following components:\par
\begin{itemize}
	\item Landscapes of the benchmark functions;
	\item Search history of the sparrow population;
	\item Exploration-exploitation ratio curves;
	\item Population diversity curves;
	\item Convergence curves.
\end{itemize}

The search history plots depict the spatial distribution of sparrow individuals during the search process. In these plots, red circles represent the global optimum, whereas blue circles indicate individual search trajectories. The results clearly show that GeoSSA is able to effectively explore the entire search space.\par
For unimodal functions (e.g., F1-F6), GeoSSA demonstrates extremely fast convergence; individuals rapidly cluster near the optimal region within a small number of iterations. For more complex functions (e.g., F7-F8, F14-F17-F23), which contain numerous local optima, GeoSSA performs rapid early-stage global exploration followed by finely tuned exploitation in later stages. The results indicate that sparrow individuals traverse most of the search space, with their trajectories ultimately concentrating near the global optimum.\par
Regarding exploration-exploitation balance, GeoSSA exhibits excellent dynamic regulation capability. For functions such as F1-F7 and F9-F13, GeoSSA maintains a high exploration ratio during early iterations and then quickly intensifies exploitation, highlighting its strong global search ability. For more challenging functions (e.g., F17-F23), GeoSSA begins with moderately high exploitation and gradually stabilizes the exploration-exploitation trend, demonstrating robust global and local search capabilities.\par
Moreover, for multimodal functions such as F8-F23, the population diversity curves show that GeoSSA maintains significant fluctuation at a relatively high level throughout the search. This indicates that GeoSSA effectively preserves population diversity, preventing premature convergence caused by population clustering. Overall, GeoSSA exhibits strong performance in exploration depth, convergence speed, and stability, achieving an excellent balance between global exploration and local exploitation.

\subsection{Comparative Experiment}
\begin{figure}[htbp]
    \centering
    \includegraphics[width=\textwidth]{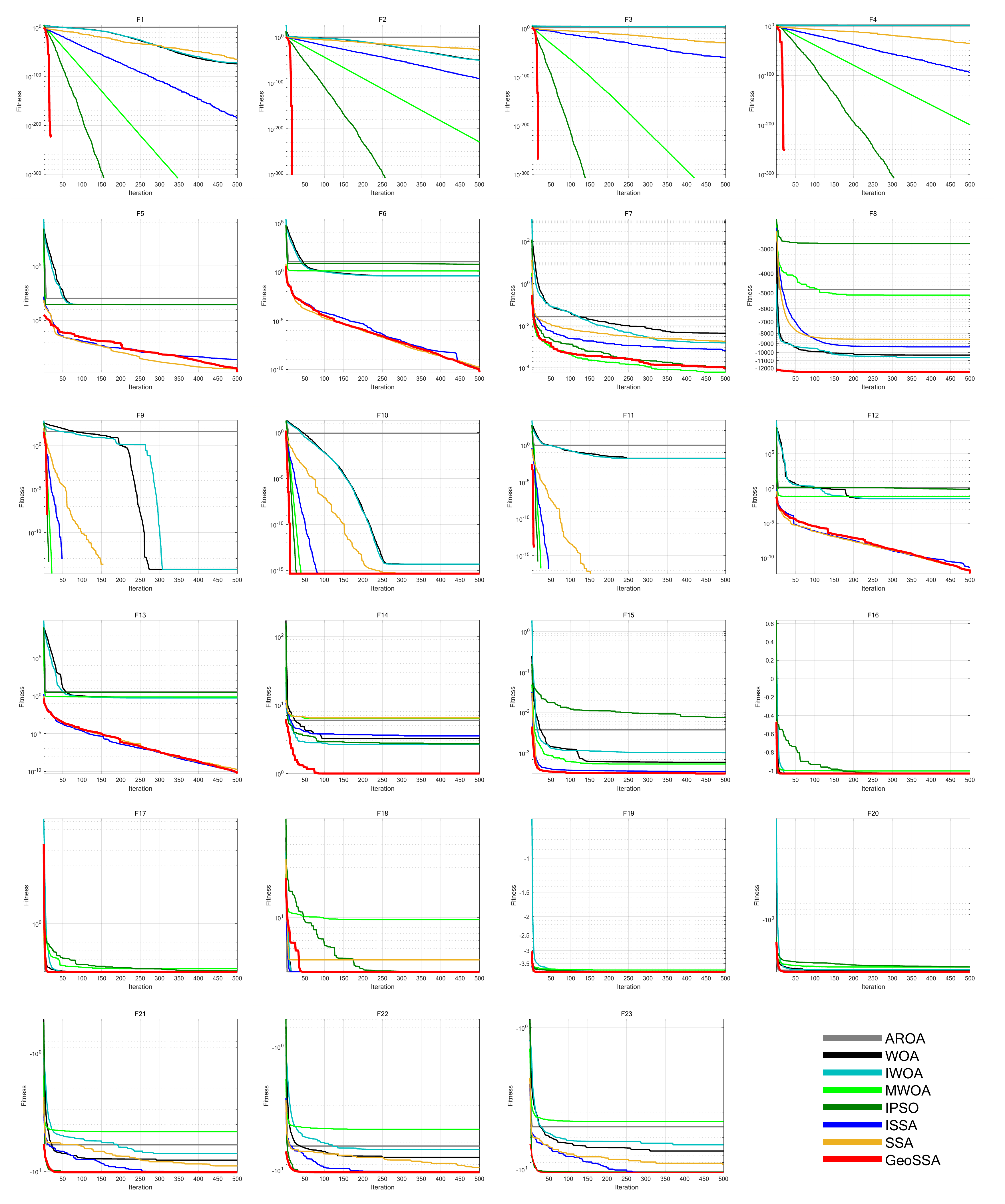}
    \caption{Iterative curves of the algorithms in the comparative experiment.}
    \label{differ1}
\end{figure}

To validate the superiority of GeoSSA, it was compared with the Attraction-Repulsion Optimization Algorithm (AROA) \cite{AROA}, Whale Optimization Algorithm (WOA) \cite{WOA}, IWOA \cite{IWOA}, MWOA \cite{MWOA}, IPSO \cite{IPSO}, ISSA \cite{ISSA}, and SSA on the benchmark functions listed in Table~\ref{table1}. The parameter settings for each algorithm are shown in Table~\ref{setting}. The number of iterations was set to $T$=500 and the population size to $N$=30. Each algorithm was independently executed 30 times on the 23 benchmark functions, and both parametric and nonparametric statistical tests were performed. The performance metrics include the average fitness ($Ave$), standard deviation ($Std$), $p$-values from the Wilcoxon signed-rank test, and Friedman ranks. The results are illustrated in Fig.~\ref{differ1}, Table~\ref{differ2} and Table~\ref{differ3}.\par

\begin{table}[htbp]
    \centering
    \caption{Parameter settings for different metaheuristic algorithms}
    \begin{tabular}{ccc}
        \toprule
        Algorithm & Parameter & Value \\
        \midrule
		AROA \cite{AROA} & Attraction factor $c$ & 0.95 \\ 
		& Local search scaling factor 1 & 0.15 \\ 
		& Local search scaling factor 2 & 0.6 \\ 
		& Attraction probability 1 & 0.2 \\ 
		& Local search probability & 0.8 \\ 
		& Expansion factor & 0.4 \\ 
		& Local search threshold 1 & 0.9 \\
		& Local search threshold 2 & 0.85 \\ 
		& Local search threshold 3 & 0.9 \\ 
		\midrule
		WOA \cite{WOA} & Convergence factor $a$ & 2 decrease to 0 \\
        & Spiral factor $b$ & 1 \\ 
		\midrule
		IWOA \cite{IWOA} & Convergence Factor $a$ & 2 decrease to 0 \\
        & Spiral Factor $b$ & 1 \\
		\midrule
		MWOA~\cite{MWOA} & Convergence Factor $a$ & 2 decreasing to 0 \\ 
        & Spiral Factor $b$ & 1 \\ 
        & ${CF}_1$ & 2.5 \\ 
        & ${CF}_2$ & 2.5 \\
        \midrule
		IPSO~\cite{IPSO} & Inertia Weight $\omega$  & 0.9 decreasing to 0 \\
            &  $C_1$ & 2 \\ 
            &  $C_2$ & 2 \\
        \midrule
		ISSA \cite{ISSA} & $ST$ & 0.7\\
		& $number of the producers$ & 0.3\\
		\midrule
		SSA \cite{SSA} & $ST$ & 0.7\\
		& $number of the producers$ & 0.3\\
		\midrule
		GeoSSA  & $ST$ & 0.7\\
		& $number of the producers$ & 0.3\\
        \bottomrule
        \label{setting}
    \end{tabular}
\end{table}

The results demonstrate that GeoSSA achieves the best overall performance among all algorithms and provides significant improvements over the original SSA. GeoSSA also shows strong competitiveness against classical metaheuristic algorithms, high-performance SSA variants, and advanced improved metaheuristics.\par
According to Fig.~\ref{differ1} and Table~\ref{differ2}, GeoSSA obtains the best average fitness ($Ave$) and standard deviation ($Std$) on most benchmark functions. GeoSSA consistently converges rapidly to the optimal solution with high accuracy, indicating strong adaptability and robustness across diverse optimization tasks. However, on F7, GeoSSA performs slightly worse than MWOA. This observation is consistent with the 'No Free Lunch Theorem', which states that no single algorithm can outperform all others on every optimization problem. Although GeoSSA achieves excellent overall performance, certain algorithms may outperform it on specific tasks, underscoring the importance of developing specialized strategies for different categories of optimization problems.\par
From the Wilcoxon signed-rank test results in Table~\ref{differ3}, GeoSSA exhibits statistically significant differences compared with AROA and IWOA on all benchmark functions. GeoSSA shows comparable performance to some algorithms on a few functions, where differences are not statistically significant. For example, GeoSSA and MWOA achieve nearly identical results on F1 and F3. Similarly, GeoSSA and IPSO perform almost identically on F1-F4, indicating that these algorithms are capable of quickly converging to the global optimum within the given iteration budget. For F9-F11, GeoSSA demonstrates no significant difference relative to MWOA, IPSO, and ISSA, reflecting convergence similarities near optimal regions. Finally, for F7, the $p$-value for MWOA exceeds 0.05, indicating that GeoSSA performs worse than MWOA on this function.\par
Based on the results of Friedman tests in Table~\ref{differ3}, GeoSSA achieves an average rank of 1.9514, the best among all algorithms. ISSA ranks second (2.9594), followed by SSA (3.3239). IPSO, IWOA, and WOA rank fourth, fifth, and sixth with average Friedman values of 5.0362, 5.1906, and 5.1978, respectively. MWOA and AROA rank seventh and eighth with scores of 5.3739 and 6.9667. These results clearly show that GeoSSA significantly outperforms other SOTA metaheuristic algorithms.

\begin{sidewaystable}[htbp]
\centering
\caption{Parametric results of the algorihms in the comparative experiment.}
\begin{tabular}{cccccccccc}
\toprule
Function & Metrics & AROA & WOA & IWOA & MWOA & IPSO & ISSA & SSA & GeoSSA \\
\midrule
F1  & Ave & 4.0514E+00 & 9.1772E-75 & 2.6908E-73 & 0.0000E+00 & 0.0000E+00 & 7.6437e-186 & 2.0231e-65 & 0.0000E+00 \\
    & Std & 2.4914E+00 & 2.8932E-74 & 1.0220E-72 & 0.0000E+00 & 0.0000E+00 & 8.4520E-186 & 1.1067E-64 & 0.0000E+00 \\
F2  & Ave & 6.4229E-01 & 1.5238E-50 & 5.7738E-51 & 1.2639E-229 & 0.0000E+00 & 2.2066E-91 & 3.2283E-29 & 0.0000E+00 \\
    & Std & 1.6465E-01 & 7.7098E-50 & 1.8475E-50 & 2.4524E-229 & 0.0000E+00 & 1.0365E-90 & 1.6139E-28 & 0.0000E+00 \\
F3  & Ave & 2.3001E+02 & 4.9269E+04 & 1.0891E+04 & 0.0000E+00 & 0.0000E+00 & 1.0357E-60 & 9.2475E-30 & 0.0000E+00 \\
    & Std & 2.6737E+02 & 1.2059E+04 & 1.5699E+04 & 0.0000E+00 & 0.0000E+00 & 5.6684E-60 & 5.0259E-29 & 0.0000E+00 \\
F4  & Ave & 1.6317E+00 & 5.0922E+01 & 1.6369E+01 & 1.0452E-200 & 0.0000E+00 & 1.4676E-94 & 4.5398E-36 & 0.0000E+00 \\
    & Std & 4.9537E-01 & 2.5849E+01 & 1.4501E+01 & 2.3521E-200 & 0.0000E+00 & 5.5654E-94 & 2.4581E-35 & 0.0000E+00 \\
F5  & Ave & 1.0455E+02 & 2.8104E+01 & 2.7792E+01 & 2.8715E+01 & 2.8932E+01 & 2.6692E-04 & 3.2808E-05 & 1.8574E-05 \\
    & Std & 9.3119E+01 & 4.7737E-01 & 4.3899E-01 & 1.3051E-01 & 9.1355E-02 & 3.6679E-04 & 6.7227E-05 & 4.1676E-05 \\
F6  & Ave & 1.0953E+01 & 4.1545E-01 & 3.7757E-01 & 1.2468E+00 & 5.8354E+00 & 1.2349E-10 & 1.6312E-10 & 5.4264E-11 \\
    & Std & 3.0083E+00 & 2.3947E-01 & 2.5550E-01 & 4.0110E-01 & 5.6576E-01 & 2.0486E-10 & 5.4238E-10 & 1.4801E-10 \\
F7  & Ave & 2.6707E-02 & 4.3968E-03 & 1.5294E-03 & 6.2096E-05 & 1.0627E-04 & 6.7477E-04 & 1.7292E-03 & 9.9504E-05 \\
    & Std & 1.7806E-02 & 3.9177E-03 & 1.6834E-03 & 6.0112E-05 & 1.1830E-04 & 4.8201E-04 & 1.4339E-03 & 1.0421E-04 \\
F8  & Ave & -4.8004E+03 & -1.0315E+04 & -1.0616E+04 & -5.1360E+03 & -2.8207E+03 & -9.3704E+03 & -8.5760E+03 & -1.2569E+04 \\
    & Std & 5.5412E+02 & 1.7152E+03 & 1.8129E+03 & 1.9636E+03 & 7.9845E+02 & 9.0770E+02 & 7.2792E+02 & 5.2754E-05 \\
F9  & Ave & 4.0975E+01 & 5.6843E-15 & 5.6843E-15 & 0.0000E+00 & 0.0000E+00 & 0.0000E+00 & 0.0000E+00 & 0.0000E+00 \\
    & Std & 5.8011E+01 & 2.2884E-14 & 1.7345E-14 & 0.0000E+00 & 0.0000E+00 & 0.0000E+00 & 0.0000E+00 & 0.0000E+00 \\
F10 & Ave & 7.8930E-01 & 4.4705E-15 & 4.7073E-15 & 4.4409E-16 & 4.4409E-16 & 4.4409E-16 & 4.4409E-16 & 4.4409E-16 \\
    & Std & 3.1087E-01 & 2.7572E-15 & 2.3603E-15 & 0.0000E+00 & 0.0000E+00 & 0.0000E+00 & 0.0000E+00 & 0.0000E+00 \\
F11 & Ave & 9.9819E-01 & 1.6368E-02 & 1.6339E-02 & 0.0000E+00 & 0.0000E+00 & 0.0000E+00 & 0.0000E+00 & 0.0000E+00 \\
    & Std & 1.0866E-01 & 5.0668E-02 & 5.2652E-02 & 0.0000E+00 & 0.0000E+00 & 0.0000E+00 & 0.0000E+00 & 0.0000E+00 \\
F12 & Ave & 1.2626E+00 & 3.4146E-02 & 3.4287E-02 & 7.1081E-02 & 7.5448E-01 & 4.5857E-12 & 1.7784E-12 & 5.8568E-13 \\
    & Std & 3.0490E-01 & 3.8896E-02 & 2.7267E-02 & 4.9211E-02 & 1.8488E-01 & 6.0585E-12 & 2.9346E-12 & 1.0975E-12 \\
F13 & Ave & 3.8194E+00 & 6.4092E-01 & 5.2446E-01 & 7.1072E-01 & 2.8750E+00 & 6.8132E-11 & 1.7852E-10 & 5.2094E-11 \\
    & Std & 4.1069E-01 & 3.2259E-01 & 3.3007E-01 & 1.9087E-01 & 1.4062E-01 & 1.5112E-10 & 6.8525E-10 & 1.4162E-10 \\
F14 & Ave & 5.9912E+00 & 3.2233E+00 & 2.6354E+00 & 6.3416E+00 & 2.7051E+00 & 3.5309E+00 & 6.4528E+00 & 9.9800E-01 \\
    & Std & 5.0892E+00 & 3.3554E+00 & 2.9515E+00 & 4.2913E+00 & 2.6083E+00 & 4.6861E+00 & 5.5831E+00 & 1.2709E-16 \\
F15 & Ave & 3.7645E-03 & 5.9536E-04 & 1.0184E-03 & 5.3728E-04 & 7.5151E-03 & 3.5019E-04 & 3.1686E-04 & 3.1304E-04 \\
    & Std & 6.1855E-03 & 2.9886E-04 & 6.8929E-04 & 1.6152E-04 & 8.7506E-03 & 2.1575E-04 & 5.1327E-05 & 3.0416E-05 \\
F16 & Ave & -1.0316E+00 & -1.0316E+00 & -1.0316E+00 & -9.8473E-01 & -1.0292E+00 & -1.0316E+00 & -1.0316E+00 & -1.0316E+00 \\
    & Std & 1.2819E-04 & 1.4022E-09 & 1.4561E-09 & 3.8021E-02 & 5.7122E-03 & 5.2156e-16 & 5.2964E-16 & 4.6467E-16 \\
F17 & Ave & 3.9885E-01 & 3.9789E-01 & 3.9790E-01 & 4.2206E-01 & 4.0478E-01 & 3.9789E-01 & 3.9789E-01 & 3.9789E-01 \\
    & Std & 3.7027E-03 & 8.1458E-06 & 8.1458E-06 & 3.1797E-02 & 8.3924E-03 & 3.2434E-16 & 0.0000E+00 & 0.0000E+00 \\
F18 & Ave & 3.0005E+00 & 3.0000E+00 & 3.0001E+00 & 9.5037E+00 & 3.0038E+00 & 3.0000E+00 & 3.9000E+00 & 3.0000E+00 \\
    & Std & 9.1161E-04 & 6.5608E-05 & 2.0970E-04 & 1.1091E+01 & 7.1699E-03 & 1.9963E-15 & 4.9295E+00 & 1.9428E-15 \\
F19 & Ave & -3.8590E+00 & -3.8513E+00 & -3.8561E+00 & -3.7812E+00 & -3.8557E+00 & -3.8628E+00 & -3.8628E+00 & -3.8628E+00 \\
    & Std & 4.7376E-03 & 1.7870E-02 & 9.8997E-03 & 7.2325E-02 & 3.7930E-03 & 2.2599E-15 & 2.3061e-15 & 2.0927E-15 \\
F20 & Ave & -3.1934E+00 & -3.2299E+00 & -3.1699E+00 & -2.9755E+00 & -2.9614E+00 & -3.2782E+00 & -3.2705E+00 & -3.3141E+00 \\
    & Std & 9.5412E-02 & 1.0302E-01 & 6.9113E-02 & 1.8864E-01 & 1.9826E-01 & 6.6370E-02 & 5.9923E-02 & 3.0164E-02 \\
F21 & Ave & -5.9550E+00 & -8.0223E+00 & -7.0654E+00 & -4.6152E+00 & -1.0123E+01 & -1.0153E+01 & -8.9637E+00 & -1.0153E+01 \\
    & Std & 2.8343E+00 & 2.6676E+00 & 2.5051E+00 & 6.5956E-01 & 2.5959E-02 & 2.8558E-10 & 2.1931E+00 & 4.7114E-15 \\
F22 & Ave & -6.3714E+00 & -7.8768E+00 & -6.8151E+00 & -4.6617E+00 & -1.0366E+01 & -1.0403E+01 & -9.5170E+00 & -1.0403E+01 \\
    & Std & 3.3719E+00 & 2.9662E+00 & 2.4909E+00 & 5.2371E-01 & 1.9359E-02 & 2.6798E-15 & 2.0147E+00 & 1.4752E-15 \\
F23 & Ave & -5.0240E+00 & -7.4556E+00 & -6.7415E+00 & -4.6137E+00 & -1.0503E+01 & -1.0536E+01 & -9.2673E+00 & -1.0536E+01 \\
    & Std & 3.0601E+00 & 3.1778E+00 & 2.5068E+00 & 3.4342E-01 & 2.0370E-02 & 6.4302E-15 & 2.3227E+00 & 2.9133E-15 \\
\bottomrule
\end{tabular}
\label{differ2}
\end{sidewaystable}

\begin{table}[htbp]
    \centering
    \caption{Non-parametric results of the algorihms in the comparative experiment.}
    \begin{tabular}{cccc}
        \toprule
        Algorithm & Rank & Average Friedman Value & +/=/- \\
        \midrule
        AROA   & 8 & 6.9667   & 23/0/0 \\
        WOA    & 6 & 5.1978   & 23/0/0 \\
		IWOA   & 5 & 5.1906   & 23/0/0 \\
        MWOA   & 7 & 5.3739   & 17/5/1 \\
        IPSO   & 4 & 5.0362   & 16/7/0 \\
        ISSA   & 2 & 2.9594   & 20/3/0 \\
        SSA    & 3 & 3.3239   & 19/4/0 \\
        GeoSSA & \textbf{1} & \textbf{1.9514}   & - \\
        \bottomrule
    \end{tabular}
    \label{differ3}
\end{table}

\subsection{Overall Effectiveness}
Table~\ref{OE} summarizes all performance results of GeoSSA and other competitors by a useful metric named overall effectiveness ($OE$). In Table~\ref{OE}, $w$ indicates win, $t$ indicates tie and $l$ indicates loss. The OE of each algorithm is computed by Eq.~\ref{OEEQ}~\cite{MDTE}.
\begin{equation}
    OE = \frac{N-L}{N} \cdot 100 \%
    \label{OEEQ}
\end{equation}
where $N$ is the total number of tests; $L$ is the total number of losing tests for each algorithm.\par
The results show that GeoSSA achieves an $OE$ value of 95.65\%, demonstrating strong competitiveness across all benchmark functions. Overall, GeoSSA is the most effective algorithm among all competitors.
\begin{table}[htbp]
    \centering
    \caption{Effectiveness of GeoSSA and other competitors.}
    \begin{tabular}{ccccccccc} 
    \toprule
    Metrics & AROA & WOA & IWOA & MWOA & IPSO & ISSA & SSA & GeoSSA \\ 
    \midrule
    $w$/$t$/$l$ & 0/0/23 & 0/0/23 & 0/0/23 & 1/5/17 & 0/7/16 & 0/3/20 & 0/4/19 & 14/8/1 \\ \midrule
    $OE$ & 0.00\% & 0.00\% & 0.00\% & 26.09\%  & 30.43\% & 13.04\% & 17.39\% & \textbf{95.65\%} \\
    \bottomrule
    \end{tabular}
    \label{OE}
\end{table}

\section{UAV Path Planning}
The core objective of 3D path planning for drones is typically to minimize the path length while successfully avoiding obstacles and satisfying the drone's dynamic constraints. To facilitate the comparison of different metaheuristic optimizers' search capabilities using a single-objective optimization algorithm, this study employs a weighted composite single-objective function $F_{tc}$, which integrates three key performance metrics into a scalar objective:\par
\begin{equation}
    F_{tc}=w_1F_{pc}+w_2F_{hc}+w_3F_{sc}
    \label{eq39}
\end{equation}
where $F_{pc}$ represents the path length cost, $F_{hc}$ denotes the height cost, and $F_{sc}$ represents the drone's dynamic constraint cost. The weight coefficients satisfy $w_i \geq 0$ and $w_1+w_2+w_3=1$.\par
In all experiments, to maintain a balance between path efficiency and flight stability, the weights are uniformly set to $w_1$=0.5, $w_2$=0.3, $w_3$=0.2. This weighted strategy allows the constraints on path length, flight height, and trajectory smoothness to be addressed within the single-objective framework, enabling comparisons among all the algorithms based on the same scalar criterion. Mathematically, the discretized path is represented by $g$ waypoints: 
\begin{equation}
    P=\{P_1,\dots,P_g\}
    \label{eq40}
\end{equation}
The path length cost term is defined as:\par
\begin{equation}
    F_{pc}=\sum_{m=1}^{g-1}\|P_{m+1}-P_m\|_2
    \label{eq41}
\end{equation}
The height cost term is defined as the standard deviation of the height samples:\par
\begin{equation}
    F_{hc}=\sqrt{\frac{1}{g}\sum_{m=1}^g(z_m-\bar{z})^2}
    \label{eq42}
\end{equation}
The dynamic constraint cost is expressed by the cumulative angle between adjacent segments:\par
\begin{equation}
    F_{sc}=\sum_{m=1}^{g-2}\arccos\left(\frac{(P_{m+1}-P_m)\cdot(P_{m+2}-P_{m+1})}{\|P_{m+1}-P_m\|\|P_{m+2}-P_{m+1}\|}\right)
    \label{eq43}
\end{equation}
The obstacle constraint is incorporated into the objective function through a penalty term (assigning a large penalty value to solutions that conflict with obstacles), leading to the final optimization problem formulation:\par
\begin{equation}
    \min_PF(P)=F_{tc}(P)+\Phi_{{obs}}(P),
    \label{eq44}
\end{equation}
where the penalty term $\Phi_{{obs}}$ uses a combination of distance threshold and squared penalties to prioritize feasible solutions.\par
\begin{figure}[htbp]
    \centering
    \includegraphics[width=\textwidth]{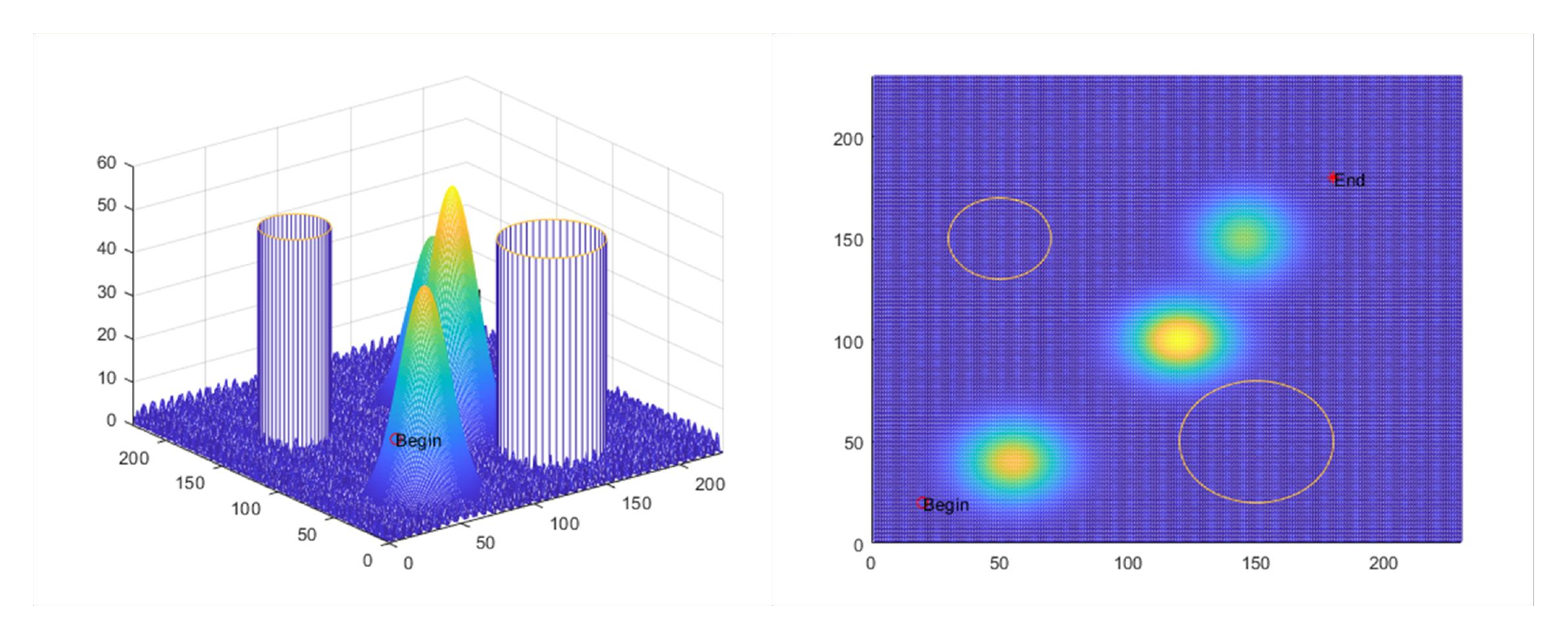}
    \caption{Terrain information of the UAV path planning problem.} 
    \label{map_info}
\end{figure}

In this study, the starting point is set as (20,20,20) and the destination as (180,180,20). Several irregular static obstacles are arranged in the 3D scene to simulate complex terrain. Fig.~\ref{map_info} is the terrain information of the UAV path planning problem. The proposed GeoSSA is compared with selected algorithms, including AROA, WOA, IWOA, MWOA, IPSO, ISSA and SSA. The number of iterations is set to $T$=500, and the population size is $N$=30. The parameter settings for each algorithm are shown in Table~\ref{setting}. To evaluate the stability and statistical performance of the algorithms, each algorithm is independently repeated 30 times. The performance metric is the $F_{tc}$ value obtained from each run, and the average fitness ($Ave$) and standard deviation ($Std$) of each algorithm are recorded to compare their optimization quality and robustness.\par

\begin{figure}[htbp]
    \centering
    \includegraphics[width=\textwidth]{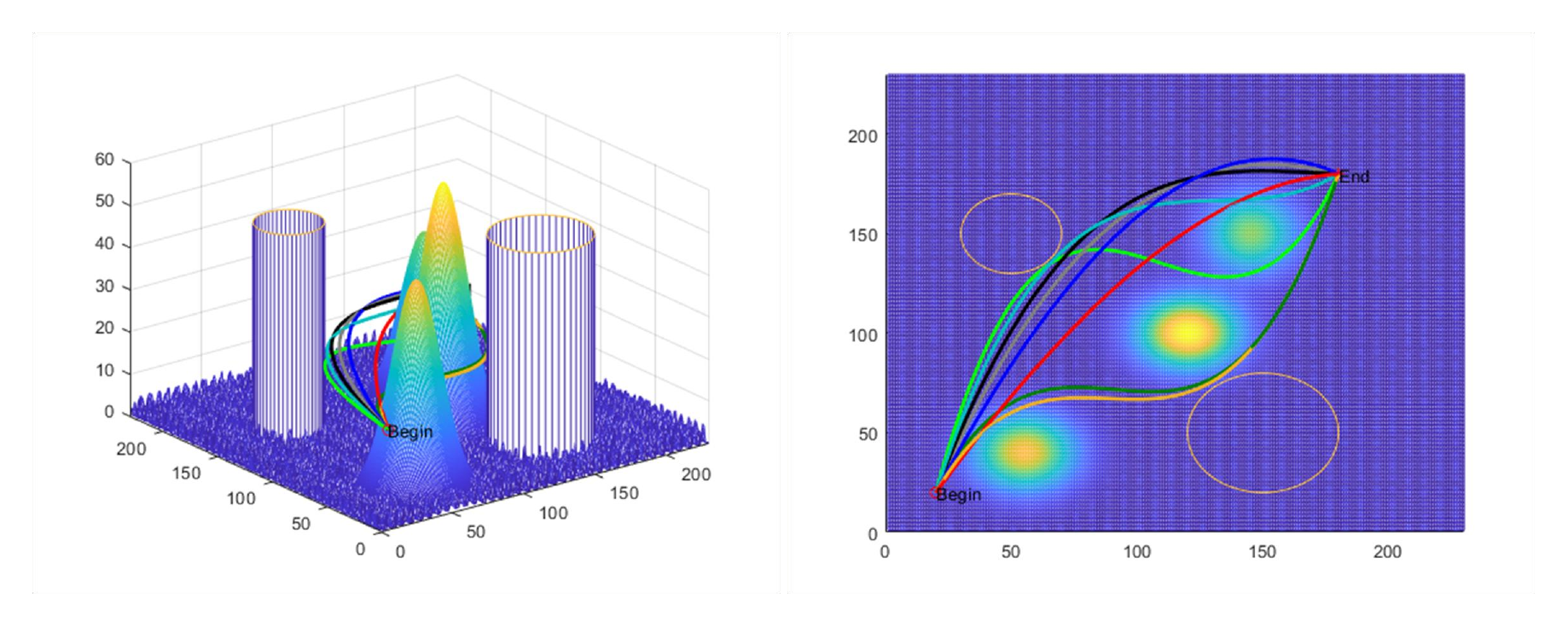}
    \caption{comparison of the paths generated by the algorithms.} 
    \label{3dpp_all}
\end{figure}

\begin{figure}[htbp]
    \centering
    \includegraphics[width=0.8\textwidth]{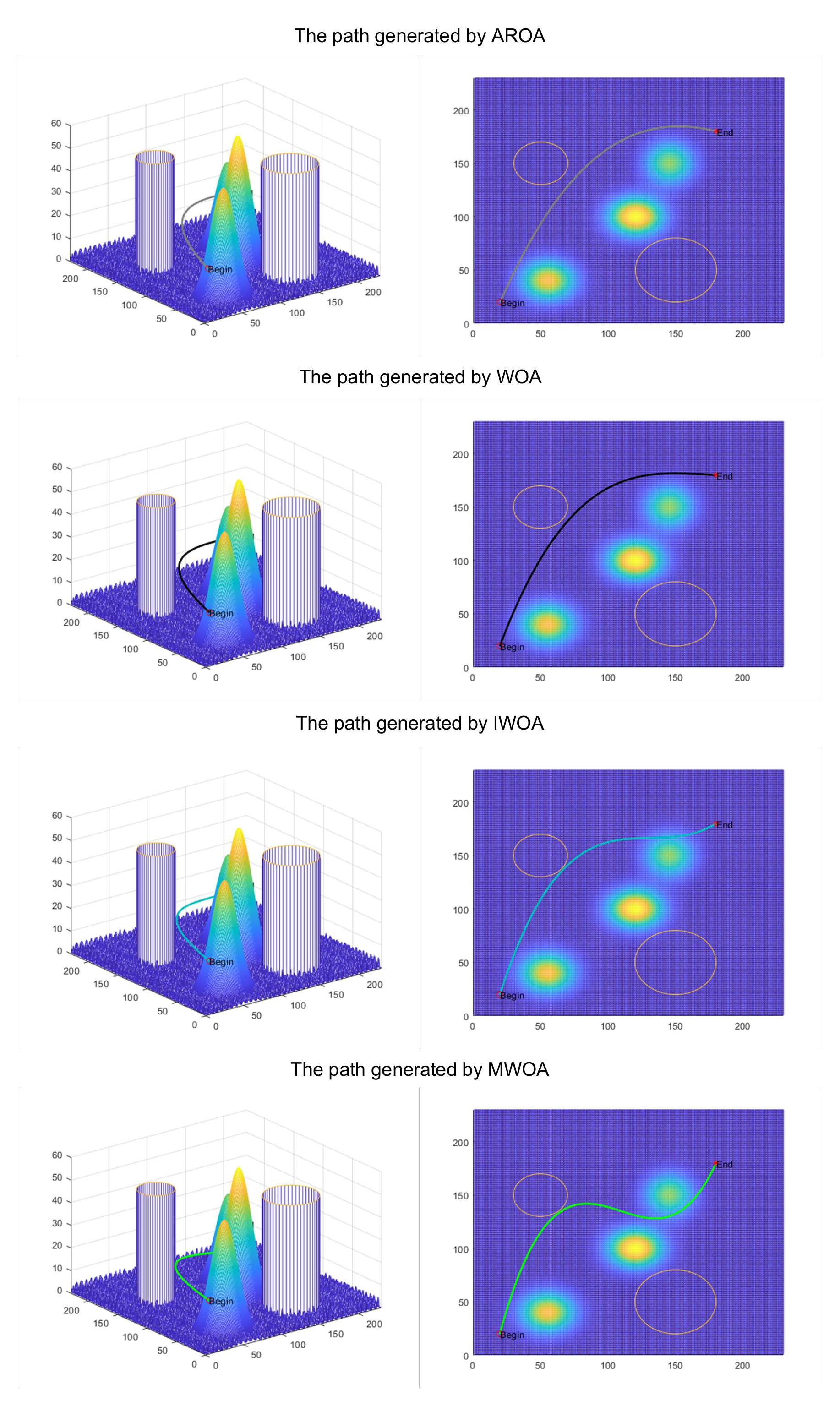}
    \caption{Paths generated by the algorithms in solving the UAV path planning problem.} 
    \label{3dpp2}
\end{figure}

\begin{figure}[htbp]
    \centering
    \includegraphics[width=0.8\textwidth]{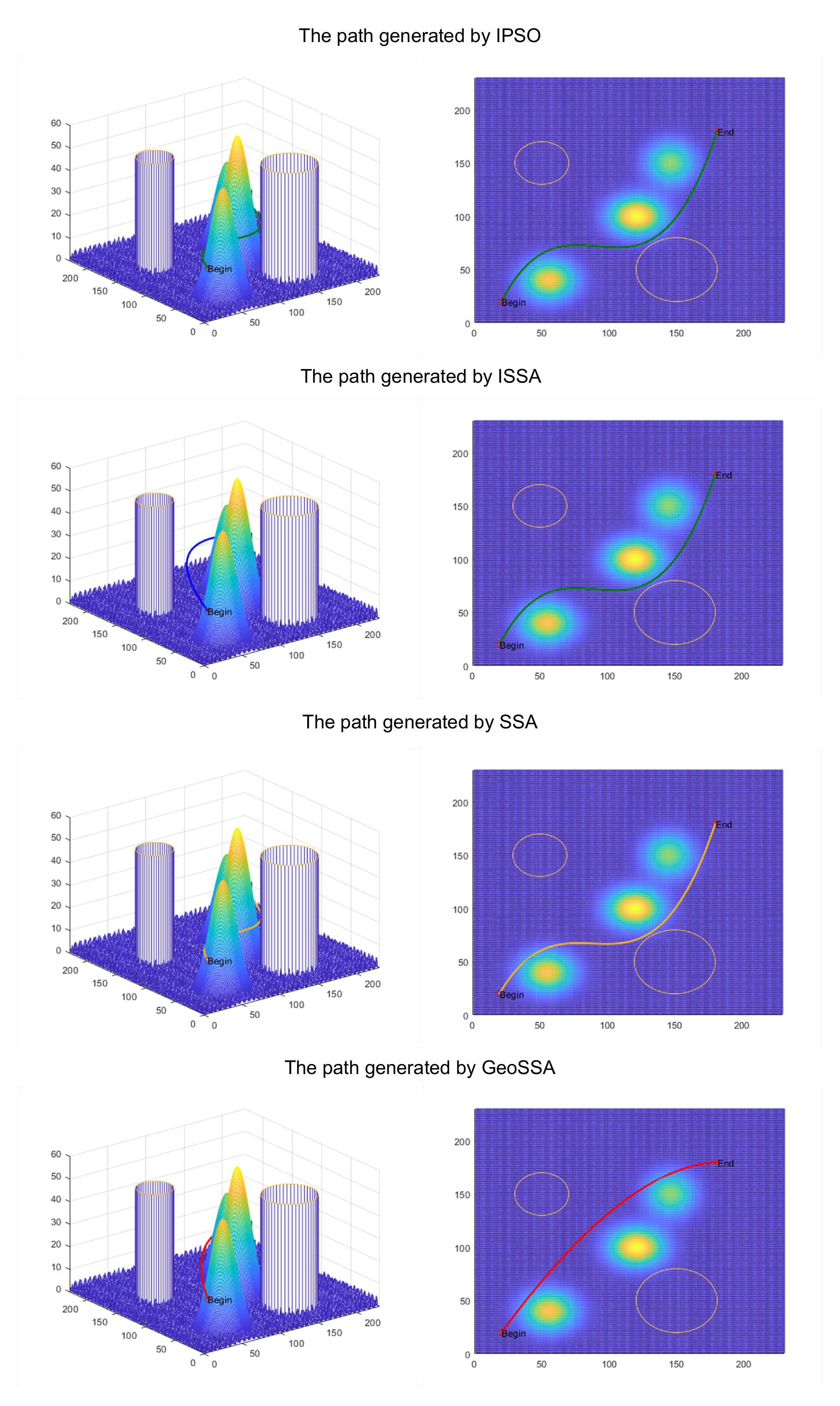}
    \caption{Paths generated by the algorithms in solving the UAV path planning problem (Continued).} 
    \label{3dpp3}
\end{figure}

\begin{figure}[htbp]
    \centering
    \includegraphics[width=0.8\textwidth]{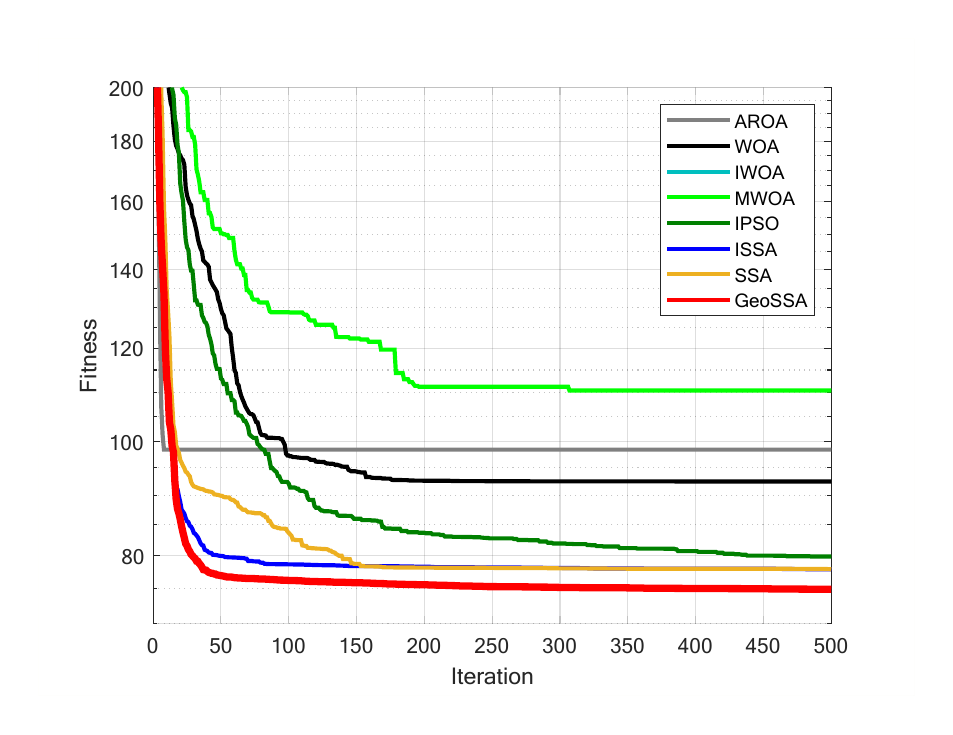}
    \caption{Iteration curves of the algorithms in solving the UAV path planning problem.} 
    \label{3dpp4}
\end{figure}

As shown in Tabel~\ref{3dpp1}, Fig.~\ref{3dpp_all}, Fig.~\ref{3dpp2}, Fig.~\ref{3dpp3} and Fig.~\ref{3dpp4}, the experimental results indicate that GeoSSA achieves the best average performance (74.8472) and the smallest standard deviation (0.95654), demonstrating that the algorithm not only attains lower objective values across multiple independent runs but also exhibits outstanding stability and robustness. In comparison, IPSO, ISSA, and SSA produce reasonably good solutions in most trials (with average values concentrated in the 77-80 range), yet their standard deviations are consistently higher than that of GeoSSA, indicating greater variability in their results. Furthermore, AROA, WOA, and MWOA show noticeably higher means and variances. Notably, IWOA presents an extremely large mean (3.33E+31) and variance (1.83E+32), suggesting that this method is prone to premature convergence or entrapment in local optima when addressing UAV path-planning tasks, resulting in poor or even unacceptable trajectory solutions.

\begin{table}[htbp]
\centering
\caption{Comparison results of the algorithms in UAV Path Planning.}
\begin{tabular}{ccc}
\toprule
\textbf{Algorithm} & \textbf{Metrics} & \textbf{Value} \\
\midrule
AROA     & Ave  & 98.3896  \\
        & Std  & 35.0485  \\
WOA    & Ave  & 92.431 \\
        & Std  & 39.8221    \\
IWOA   & Ave  & 3.33E+31  \\
        & Std  & 1.83E+32  \\
MWOA  & Ave  & 110.4649 \\
        & Std  & 22.6079 \\
IPSO    & Ave  & 79.7978  \\
        & Std  & 1.9662    \\
ISSA    & Ave  & 77.8496  \\
        & Std  & 21.2555  \\
SSA    & Ave  & 77.8214  \\
        & Std  & 11.2765   \\
GeoSSA   & Ave  & \textbf{74.8472}  \\
        & Std  & \textbf{0.95654}   \\
		\midrule
\end{tabular}
\label{3dpp1}
\end{table}

Additional analyses based on convergence curves and trajectory visualizations further confirm that GeoSSA exhibits a rapid early-stage convergence rate while remaining stable in later iterations, indicating a well-balanced trade-off between exploration and exploitation under the weighted objective. The trajectory plots also show that GeoSSA produces shorter flight paths with smoother turns, successfully avoids obstacles, and simultaneously satisfies both dynamic and altitude constraints. Overall, GeoSSA demonstrates superior optimization capability and the highest stability in solving UAV path planning problems.

\section{Engineering Design Optimization}
Engineering design optimization aims to identify design solutions that achieve optimal performance, minimal cost, or the most rational structural configuration for practical engineering applications while satisfying specified constraints. Because engineering design problems frequently involve complex nonlinear constraints, coexisting discrete and continuous variables, multimodal landscapes, and nondifferentiable objective functions, classical analytical and gradient-based optimization methods face substantial limitations when solving such problems. Consequently, meta-heuristic algorithms—characterized by strong global search capability, no requirement for gradient information, and high robustness—are widely employed for the analysis and solution of these problems.\par
In this study, constraint handling in all engineering design optimization problems is performed using the Penalty Function Method, which is one of the most widely adopted and effective techniques in meta-heuristic optimization. Specifically, constrained optimization problems are transformed into unconstrained ones by augmenting the original objective function with penalty terms corresponding to constraint violations. When a candidate solution violates one or more constraints, a sufficiently large penalty is imposed, thereby discouraging infeasible solutions and guiding the search process toward the feasible region. The penalty coefficients are selected as fixed large constants, following common practice in the related literature, to ensure that feasibility is strictly prioritized over objective improvement without introducing excessive numerical instability.\par
To comprehensively evaluate the applicability and advantages of the proposed GeoSSA algorithm in realistic engineering design scenarios, this section selects four representative engineering design optimization problems for testing: Corrugated Bulkhead design problem, Piston Lever design problem, Reactor Network design problem, and an Industrial Refrigeration System design problem. These problems span mechanical structures, chemical process design, and industrial equipment, each exhibiting strong constraints, high coupling, and pronounced nonlinearity, thereby providing a thorough assessment of algorithm stability and solution capability across diverse practical settings. The experiments in this study were conducted on a system equipped with Windows 11 (64-bit), an Intel(R) Core(TM) i5-8300H CPU @ 2.30 GHz processor, 8 GB RAM, and MATLAB R2023a as the simulation platform. GeoSSA is compared against AROA, WOA, IWOA, MWOA, IPSO, ISSA, and SSA. The parameter settings for all algorithms are listed in Table~\ref{setting}. All algorithms were run with a unified configuration of $T = 500$ iterations and population size $N = 30$. Each algorithm was independently executed 30 times on every engineering design problem, and the average fitness ($Ave$) and standard deviation ($Std$) were recorded for performance evaluation.
The experimental results are presented in Fig.~\ref{engineering12}, Fig.~\ref{engineering22}, Fig.~\ref{engineering32}, Fig.~\ref{engineering42} and Table~\ref{engineeringmetrics}. The following subsections describe the mathematical models and constraints for each engineering design problem in detail and provide a thorough analysis of the solution performance of GeoSSA relative to the comparison algorithms.

\subsection{Corrugated Bulkhead}
\begin{figure}[htbp]
    \centering
    \includegraphics[width=0.8\textwidth]{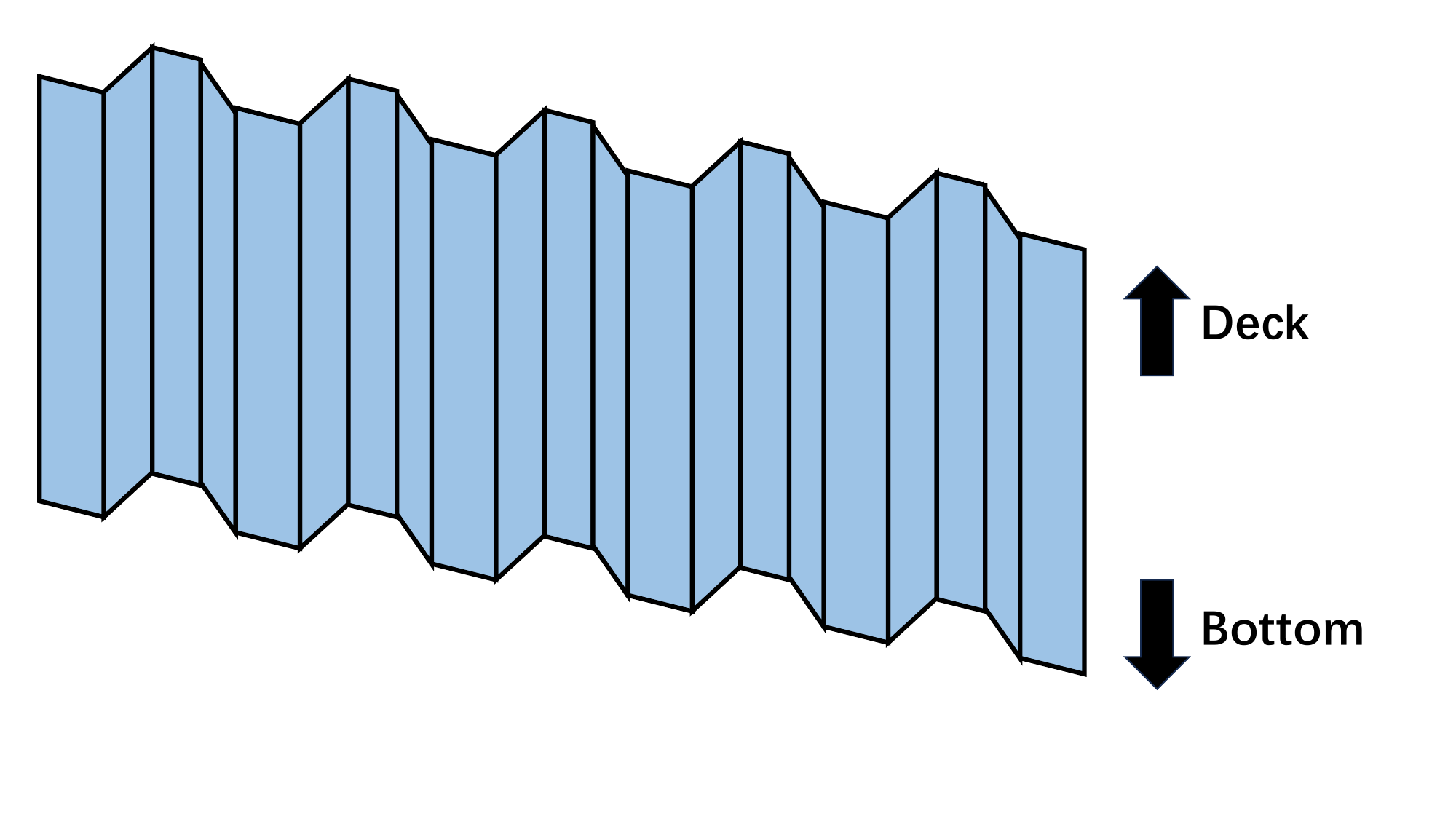}
    \caption{The structure of a corrugated bulkhead} 
    \label{engineering11}
\end{figure}

The design of corrugated bulkheads plays a crucial role in chemical plant design, particularly in the design of chemical storage tanks, reactors, pressure vessels, and other equipment. Corrugated bulkheads are typically used in containers that must withstand high pressure during storage, transportation, and chemical reactions, and are primarily intended to enhance the structural strength and stability, as shown in Fig.~\ref{engineering11}. Compared to traditional flat bulkheads, in the chemical plant environment, which is often exposed to numerous vibration sources, corrugated bulkheads increase the load-bearing capacity by distributing the special corrugated shape, thus reducing the risk of structural fatigue, deformation, and rupture. By optimizing the structure of the corrugated bulkhead, the safety, strength, durability, and cost-effectiveness of the equipment can be further improved, which in turn enhances the overall design efficiency of the chemical plant and ensures the stability and safety of the production process. The focus of the corrugated bulkhead design task is to determine the optimal shape that maximizes compressive strength. Through parameter optimization, including the length, width, height, and material thickness of the corrugation, metaheuristic algorithms seek a design that achieves maximum compressive strength with minimal material usage. This task involves four variables: length $x_1$, width $x_2$, height $x_3$, and thickness $x_4$ of the corrugations. There are six inequality constraints to ensure physical properties suitable for real-world use, such as maximum material strength, allowable deformation, and manufacturing limitations. The optimized corrugated bulkhead design achieves improved material utilization, lower production costs, and enhanced structural performance, meeting safety and durability standards. The mathematical modeling of Corrugated Bulkhead (CB) design problem is as follows:\par

\begin{flushleft}
    \textit{Variable:}
\end{flushleft}
\[
    \begin{aligned}
        x = [x_1, x_2, x_3, x_4]
    \end{aligned}
\]

\begin{flushleft}
    \textit{Minimize:}
\end{flushleft}
\begin{flushleft}
    \[
        \begin{aligned}
            f(x) &= \frac{5.885 \cdot x_4 \cdot (x_1 + x_3)}{x_1 + \sqrt{|x_3^2 - x_2^2|}} + punishment
        \end{aligned}
    \]
\end{flushleft}
\begin{flushleft}
    \textit{Subject to:}
\end{flushleft}

\begin{equation}
            g_1 = -x_4 \cdot x_2 \cdot (0.4 \cdot x_1 + \frac{x_3}{6}) + 8.94 \cdot (x_1 + \sqrt{|x_3^2 - x_2^2|}) \leq 0;
\end{equation}

\begin{equation}
            g_2 = -x_4 \cdot x_2^2 \cdot (0.2 \cdot x_1 + \frac{x_3}{12}) + 2.2 \cdot \left(8.94 \cdot (x_1 + \sqrt{|x_3^2 - x_2^2|})\right)^{\frac{4}{3}} \leq 0;
\end{equation}

\begin{equation}
            g_3 = -x_4 + 0.0156 \cdot x_1 + 0.15 \leq 0;
\end{equation}

\begin{equation}
             g_4 = -x_4 + 0.0156 \cdot x_3 + 0.15 \leq 0;
\end{equation}

\begin{equation}
            g_5 = -x_4 + 1.05 \leq 0;
\end{equation}

\begin{equation}
           g_6 = -x_3 + x_2 \leq 0;
\end{equation}

\begin{flushleft}
    \textit{Variable range:}
\end{flushleft}
\vspace{-\baselineskip}
\begin{flushleft}
    \[
    1\leq x_1 \leq100; \quad 1\leq x_2 \leq 100;  
    \]
\end{flushleft}
\vspace{-\baselineskip}
\begin{flushleft}
    \[
     1\leq x_3 \leq 100; \quad 1\leq x_4 \leq 5; 
    \]
\end{flushleft}

\begin{flushleft}
\textit{Where:}
\end{flushleft}
\begin{flushleft}
    \[
        \begin{aligned}
        punishment=10^3\cdotp\sum_{i=1}^6\max{(0,g_i)^2}
        \end{aligned}
\]
\end{flushleft}

\begin{figure}[htbp]
    \centering
    \includegraphics[width=0.8\textwidth]{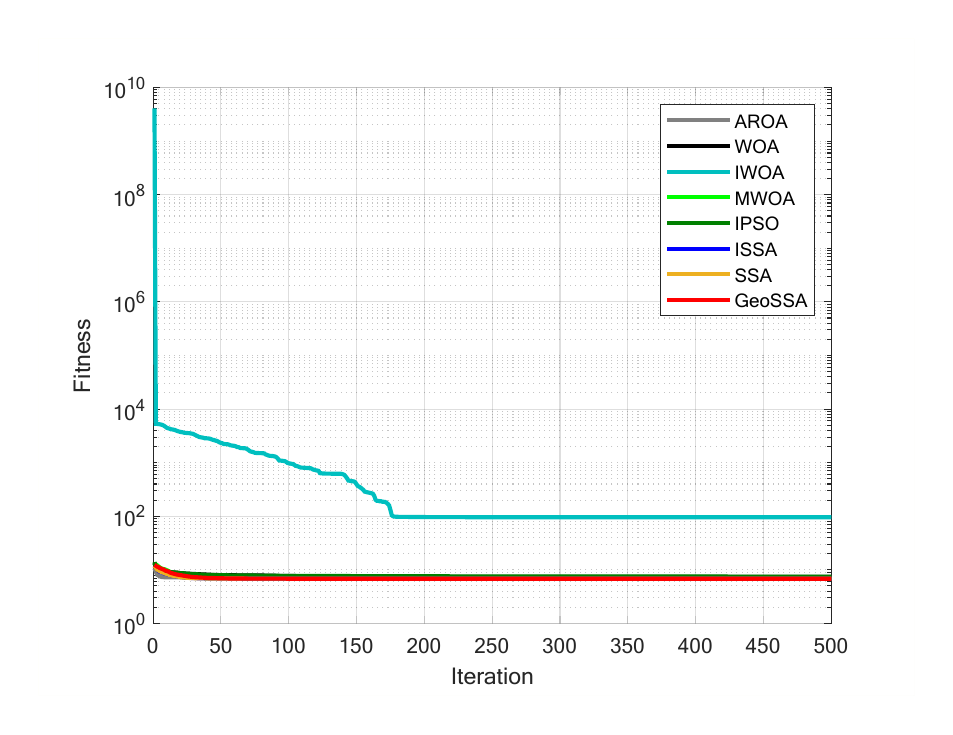}
    \caption{Iteration curves of the algorithms in solving the Corrugated Bulkhead design problem.} 
    \label{engineering12}
\end{figure}

The experimental results are presented in Fig.~\ref{engineering12} and Table~\ref{engineeringmetrics}. As shown in Table~\ref{engineeringmetrics}, in the Corrugated Bulkhead design problem, GeoSSA exhibits significantly higher stability than all other algorithms, and its optimization accuracy is the best among the compared methods. This demonstrates that GeoSSA possesses substantial advantages when addressing this type of engineering optimization problem.

\subsection{Piston Lever}
\begin{figure}[htbp]
    \centering
    \includegraphics[width=0.6\textwidth]{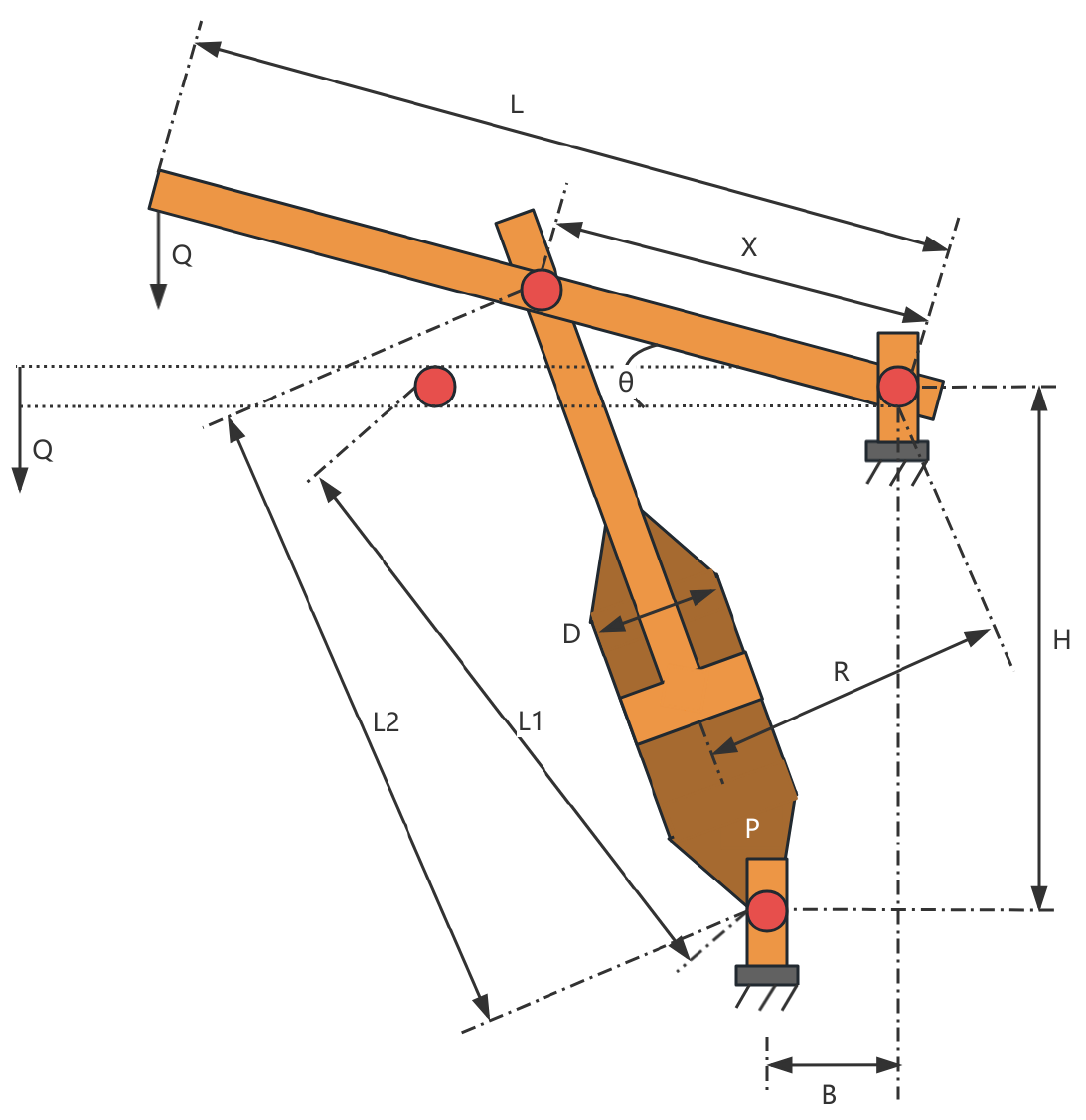}
    \caption{The structure of a piston lever \cite{lsewoa}.} 
    \label{engineering21}
\end{figure}

Piston levers are critical components in various important pieces of equipment, such as pumps and compressors, where their performance directly impacts the reliability and efficiency of the entire system. The Piston Lever (PL) design problem focuses on determining the optimal piston dimensions and material selection to ensure maximum performance in transmission systems. This design influences the efficiency, stability, and durability of mechanical systems. The optimization task seeks to balance dimensions, weight, material choice, and manufacturing cost in order to find the optimal piston size and material that maximize system efficiency and cost-effectiveness. Therefore, the Piston Lever design problem is of paramount importance in chemical plant design, and optimizing these key components is essential to ensuring the optimal performance of the system. As shown in Fig.~\ref{engineering21}, the optimization task involves four variables: piston length $x_1$, piston diameter $x_2$, material property $x_3$, and transmission rod length $x_4$, which affect the system's mechanical properties, dynamic performance, and cost.\par
The Piston Lever design problem can be described as:\par
\begin{flushleft}
    \textit{Variable:}
\end{flushleft}
\vspace{-\baselineskip}
\begin{flushleft}
    \[
    x = [x_1, x_2, x_3, x_4]
    \]
\end{flushleft}

 \begin{flushleft}
    \textit{Minimize:}
\end{flushleft}
\begin{equation}
    f(x) = 0.25\pi x_3^2(L_2-L_1)+punishment
\end{equation}

\begin{flushleft}
    \textit{Subject to:}
\end{flushleft}
\begin{equation}
    g_1=QL\cos\theta-RF \leq 0;
\end{equation}

\begin{equation}
        g_2=Q(L-x_4)-M_{\max} \leq 0;
\end{equation}

\begin{equation}
    g_3=1.2(L_2-L_1)-L_1 \leq 0;
\end{equation}

\begin{equation}
    g_4=\frac{x_3}2-x_2 \leq 0;
\end{equation}

\begin{flushleft}
    \textit{Where:}
\end{flushleft}
\vspace{-\baselineskip}
\begin{flushleft}
    \[
    Q=10000;\quad P=1500; \quad L=240; \quad F=0.25\pi Px_3^2;
    \]
\end{flushleft}
\begin{flushleft}
    \[
     M_{\max}=1.8\times10^6; \quad L_1=\sqrt{(x_4-x_2)^2+x_1^2};
    \]
\end{flushleft}
\begin{flushleft}
    \[
     L_2=\sqrt{(x_4\sin\theta+x_1)^2+(x_2-x_4\cos\theta)^2};
    \]
\end{flushleft}
\begin{flushleft}
    \[
    R=\frac{|-x_4(x_4\sin\theta+x_1)+x_1(x_2-x_4\cos\theta)|}{\sqrt{(x_4-x_2)^2+x_1^2}};
    \]
\end{flushleft}
\begin{flushleft}
    \[
        \begin{aligned}
        punishment=10^3\cdotp\sum_{i=1}^4\max{(0,g_i)^2}
        \end{aligned}
\]
\end{flushleft}

\begin{flushleft}
    \textit{Variable range:}
\end{flushleft}
\vspace{-\baselineskip}
\begin{flushleft}
    \[
    0.05\leq x_1 \leq500; \quad 0.05\leq x_2 \leq 500;  
    \]
\end{flushleft}
\vspace{-\baselineskip}
\begin{flushleft}
    \[
     0.05\leq x_3 \leq 120; \quad 0.05\leq x_4 \leq 500; 
    \]
\end{flushleft}

\begin{figure}[htbp]
    \centering
    \includegraphics[width=0.8\textwidth]{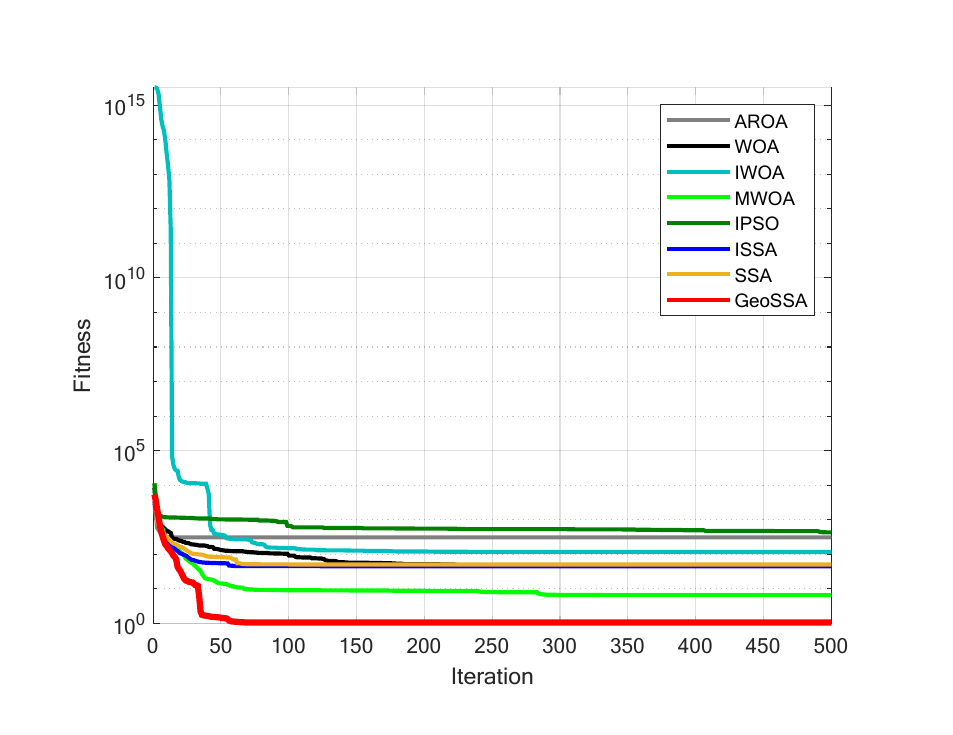}
    \caption{Iteration curves of the algorithms in solving the Piston Lever design problem.} 
    \label{engineering22}
\end{figure}

The experimental results are presented in Fig.~\ref{engineering22} and Table~\ref{engineeringmetrics}. As shown in Table~\ref{engineeringmetrics}, in the Piston Lever design problem, GeoSSA exhibits significantly higher stability than all other algorithms, and its optimization accuracy is the best among the compared methods. This demonstrates that GeoSSA possesses substantial advantages when addressing this type of engineering optimization problem.

\subsection{Reactor Network}
The Reactor Network (RN) design problem aims to optimize the configuration of chemical reactors within a chemical plant to achieve more efficient reaction processes. This involves selecting reactor types, determining their arrangement, and allocating fluid flow rates, with the objective of maximizing product concentration. By adjusting reactor configurations and operating conditions, the reaction efficiency can be improved, energy consumption reduced, and product quality enhanced.\par
\begin{figure}[htbp]
    \centering
    \includegraphics[width=0.7\textwidth]{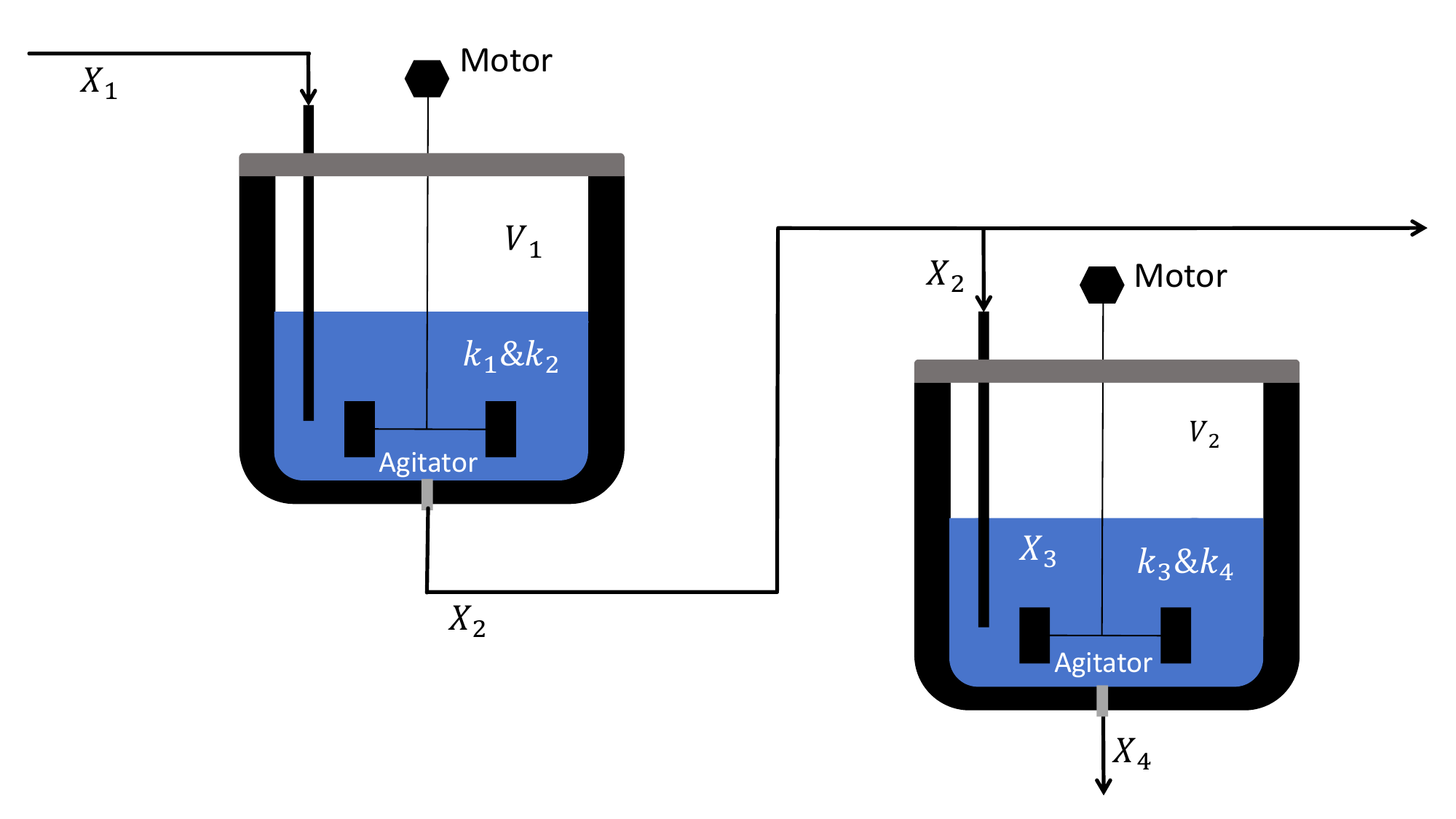}
    \caption{The structure of a reactor network} 
    \label{engineering31}
\end{figure}

As shown in Fig.~\ref{engineering31}, the problem involves four variables representing the concentrations at different reaction stages: $x_1$ denotes the reactant concentration in the first reactor, $x_2$ the product concentration in the first reactor, $x_3$ the reactant concentration in the second reactor, and $x_4$ the final product concentration. The constraint conditions $h_1$ to $h_4$ are defined as follows: $h_1$ represents the balance between reactants and products in the first reactor; $h_2$ enforces mass conservation between the first and second reactors; $h_3$ maintains the concentration balance of reactants between reactors; and $h_4$ ensures mass conservation between intermediate and final products.\par
The mathematical model of the reactor network design problem is given as follows.\par
\begin{flushleft}
    \textit{Variable:}
\end{flushleft}
\begin{flushleft}
    \[
        \begin{aligned}
            x &= [x_1, x_2, x_3, x_4, x_5, x_6]
        \end{aligned}
    \]
\end{flushleft}

\begin{flushleft}
    \textit{Minimize:}
\end{flushleft}
\begin{equation}
        f(x) = x_4+punishment
\end{equation}

\begin{flushleft}
    \textit{Subject to:}
\end{flushleft}
\begin{equation}
        g = \sqrt{x_5} + \sqrt{x_6} - 4 \leq 0;
\end{equation}
\begin{equation}
        h_1(x) = x_1 + k_1 \cdot x_2 \cdot x_5 - 1=0; 
\end{equation}
\begin{equation}
        h_2 = x_2 - x_1 + k_2 \cdot x_2 \cdot x_6=0;
\end{equation}
\begin{equation}
        h_3(x) = x_3 + x_1 + k_3 \cdot x_3 \cdot x_5 - 1=0; 
\end{equation}
\begin{equation}
        h_4 = x_4 - x_3 + x_2 - x_1 + k_4 \cdot x_4 \cdot x_6=0; 
\end{equation}

\begin{flushleft}
    \textit{Where:}
\end{flushleft}
\vspace{-\baselineskip}
\begin{flushleft}
    \[
        k_1 = 0.09755988; \quad k_2 = 0.99 \cdot k_1; \quad  k_3 = 0.0391908; 
    \]
\end{flushleft}
\vspace{-\baselineskip}
\begin{flushleft}
    \[
        k_4 = 0.9 \cdot k_3;
    \]
\end{flushleft}
\begin{flushleft}
    \[
        \begin{aligned}
        punishment=10^3\cdot \max{(0,g)^2}
        \end{aligned}
\]
\end{flushleft}

\begin{flushleft}
    \textit{Variable range:}
\end{flushleft}
\vspace{-\baselineskip}
\begin{flushleft}
    \[
        0.00001 \leq x_1 \leq 1; \quad 0.00001 \leq x_2 \leq 1; \quad 0.00001 \leq x_3 \leq 1; 
    \]
\end{flushleft}
\vspace{-\baselineskip}
\begin{flushleft}
    \[
         0.00001 \leq x_4 \leq 1; \quad 0.00001 \leq x_5 \leq 16; 
    \]
\end{flushleft}
\vspace{-\baselineskip}
\begin{flushleft}
    \[
          0.00001 \leq x_6 \leq 16;
    \]
\end{flushleft}

\begin{figure}[htbp]
    \centering
    \includegraphics[width=0.8\textwidth]{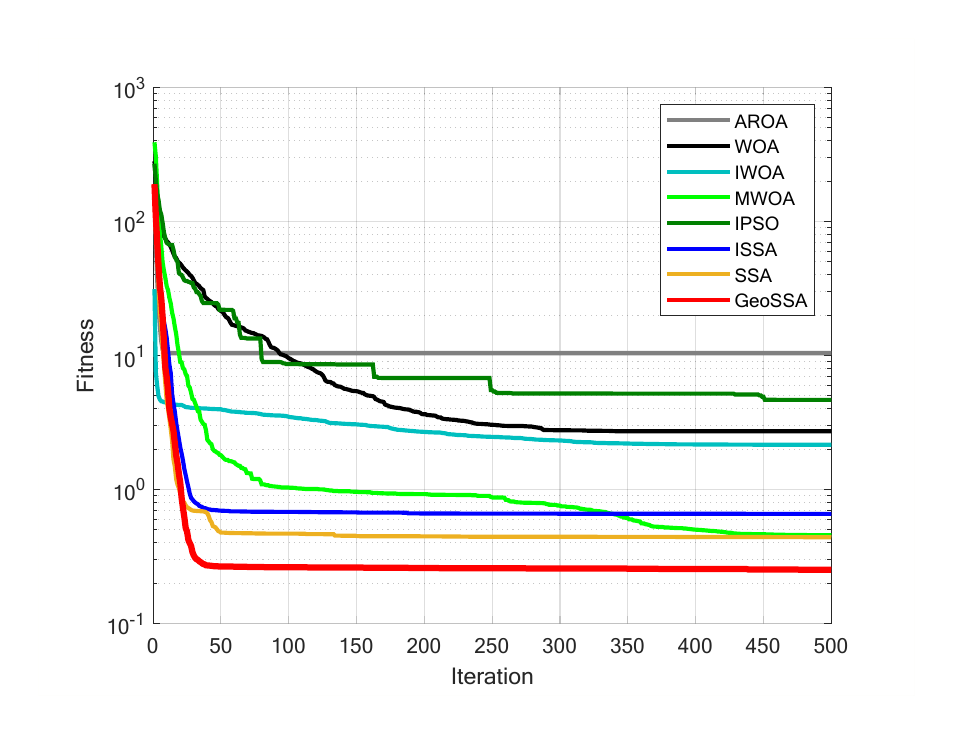}
    \caption{Iteration curves of the algorithms in solving the Reactor Network design problem.} 
    \label{engineering32}
\end{figure}

The experimental results are presented in Fig.~\ref{engineering32} and Table~\ref{engineeringmetrics}. As shown in Table~\ref{engineeringmetrics}, in the Reactor Network design problem, GeoSSA exhibits significantly higher stability than all other algorithms, and its optimization accuracy is the best among the compared methods. This demonstrates that GeoSSA possesses substantial advantages when addressing this type of engineering optimization problem.

\subsection{Industrial Refrigeration System}
In chemical industries, the industrial refrigeration system is one of the key auxiliary facilities and is widely used in various stages of chemical production. Temperature control and heat exchange are essential in operations such as chemical reactions, storage, transportation, and refining. Chemical plants typically require substantial cooling and temperature regulation to maintain reaction stability, ensure product quality, reduce energy consumption and emissions, and guarantee the safe and reliable operation of equipment.\par
\begin{figure}[htbp]
    \centering
    \includegraphics[width=\textwidth]{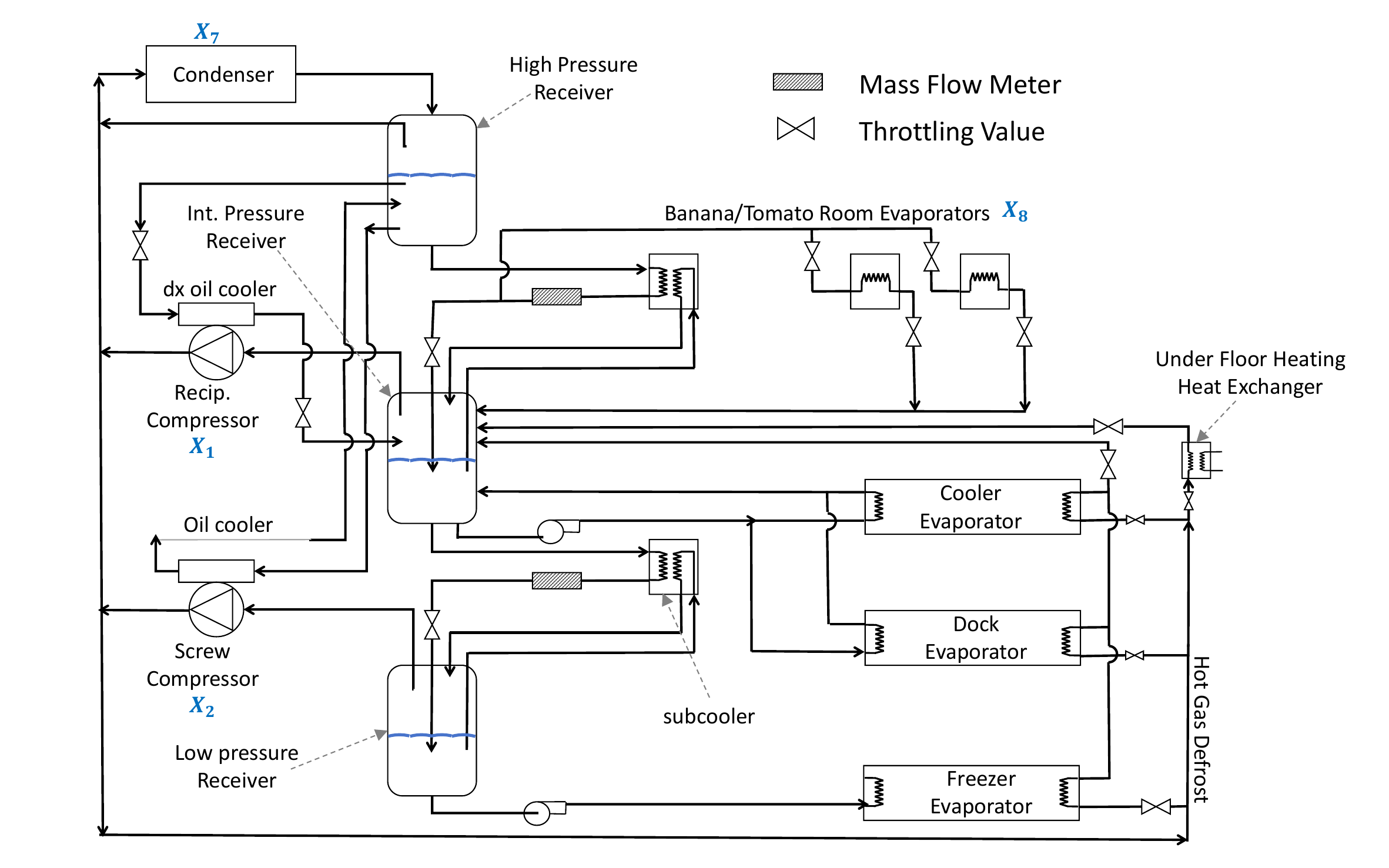}
    \caption{The structure of an industrial refrigeration system \cite{lswoa} \cite{rwoa}.} 
    \label{engineering41}
\end{figure}

The Industrial Refrigeration System (IRS) design problem aims to minimize energy consumption and operating cost while ensuring efficient cooling performance, as illustrated in Fig.~\ref{engineering41}. The objective is to configure system components—such as compressors, condensers, and evaporators—to achieve the lowest operational cost and optimal heat-exchange efficiency.\par
This problem involves fourteen decision variables: compressor power $x_1$ and 
$x_2$; refrigerant flow rates and mass flow rates $x_3$ to $x_6$; condenser and evaporator characteristics $x_7$ and $x_8$; compression ratios $x_9$ and $x_10$; temperature-related parameters $x_11$ and $x_12$; and flow parameters 
$x_13$ and $x_14$. Specifically, compressor power $x_1$ and $x_2$ determine the cooling capacity; refrigerant flow and mass flow variables $x_3$-$x_6$ describe refrigerant movement through the condenser, evaporator, and receiver; $x_7$ and $x_8$ represent condenser and evaporator sizing parameters; $x_9$ and $x_10$ define the compression ratio and compressor efficiency; $x_11$ and $x_12$ govern temperature differences for heat exchange; and $x_13$ and $x_14$ regulate cooling-water or refrigerant flow rates, thereby influencing the overall system performance.\par
The mathematical formulation of the industrial refrigeration system design problem is presented as follows.\par
\begin{flushleft}
    \textit{Variable:}
\end{flushleft}
\vspace{-\baselineskip}
\begin{flushleft}
    \[
        \begin{aligned}
            x &= [x_1, x_2, x_3, x_4, x_5, x_6, x_7, x_8, x_9, x_{10}, x_{11}, x_{12}, x_{13}, x_{14}]
        \end{aligned}
    \]
\end{flushleft}

\begin{flushleft}
    \textit{Minimize:}
\end{flushleft}
\vspace{-\baselineskip}
\begin{equation}
        y = f(x)+punishment
\end{equation}

\begin{flushleft}
    \textit{Subject to:}
\end{flushleft}
\begin{equation}
    g_1 = \frac{1.524}{x_7} - 1 \leq 0;
\end{equation}

\begin{equation}
    g_2 = \frac{1.524}{x_8} - 1 \leq 0;
\end{equation}

\begin{equation}
    g_3 = 0.07789 \cdot x_1 - \frac{2 \cdot x_9}{x_7} - 1 \leq 0;
\end{equation}

\begin{equation}
    g_4 = \frac{7.05305 \cdot x_1^2 \cdot x_{10}}{x_9 \cdot x_8 \cdot x_2 \cdot x_{14}} - 1 \leq 0;
\end{equation}

\begin{equation}
    g_5 = \frac{0.0833 \cdot x_{14}}{x_{13}} - 1 \leq 0;
\end{equation}

\begin{equation}
\begin{aligned}
g_6 =\;& \frac{47.136\,x_2^{0.333}\,x_{12}}{x_{10}}
      - 1.333\,x_8\,x_{13}^{2.1195} \\
     & + \frac{62.08\,x_{13}^{2.1195}\,x_8^{0.2}}{x_{12}\,x_{10}}
      - 1 \leq 0;
\end{aligned}
\end{equation}

\begin{equation}
    g_7 = 0.04771 \cdot x_{10} \cdot x_8^{1.8812} \cdot x_{12}^{0.3424} - 1 \leq 0;
\end{equation}

\begin{equation}
    g_8 = 0.0488 \cdot x_9 \cdot x_7^{1.893} \cdot x_{11}^{0.316} - 1 \leq 0;
\end{equation}

\begin{equation}
    g_9 = \frac{0.0099 \cdot x_1}{x_3} - 1 \leq 0;
\end{equation}

\begin{equation}
    g_{10} = \frac{0.0193 \cdot x_2}{x_4} - 1 \leq 0;
\end{equation}

\begin{equation}
    g_{11} = \frac{0.0298 \cdot x_1}{x_5} - 1 \leq 0;
\end{equation}

\begin{equation}
    g_{12} = \frac{0.056 \cdot x_2}{x_6} - 1 \leq 0;
\end{equation}

\begin{equation}
    g_{13} = \frac{2}{x_9} - 1 \leq 0;
\end{equation}

\begin{equation}
    g_{14} = \frac{2}{x_{10}} - 1 \leq 0;
\end{equation}

\begin{equation}
    g_{15} = \frac{x_{12}}{x_{11}} - 1 \leq 0;
\end{equation}

\begin{flushleft}
    \textit{Where:}
\end{flushleft}

\begin{equation}
\begin{aligned}
f(x) =\;& 63098.88\,x_2 x_4 x_{12} 
+ 5441.5\,x_2^2 x_{12} \\
&+ 115055.5\,x_2^{1.664} x_6 
+ 6172.27\,x_2^2 x_6 \\
&+ 63098.88\,x_1 x_3 x_{11} 
+ 5441.5\,x_1^2 x_{11} \\
&+ 115055.5\,x_1^{1.664} x_5 
+ 6172.27\,x_1^2 x_5 \\
& + 140.53\,x_1 x_{11} 
+ 281.29\,x_3 x_{11} \\
&+ 70.26\,x_1^2 
+ 281.29\,x_1 x_3 
+ 281.29\,x_3^2 \\
& + 14437\,x_8^{1.8812} x_{12}^{0.3424} x_{10} x_1^2 
    \frac{x_7}{x_{14} x_9}\\
& + 20470.2\,x_7^{2.893} x_{11}^{0.316} x_{12}
\end{aligned}
\end{equation}

\begin{flushleft}
    \[
        \begin{aligned}
        punishment=10^3\cdotp\sum_{i=1}^{15}\max{(0,g_i)^2}
        \end{aligned}
\]
\end{flushleft}

\begin{flushleft}
    \textit{Variable range:}
\end{flushleft}
\vspace{-\baselineskip}
\begin{flushleft}
    \[
        0.001 < x_1 < 5; \quad 0.001 < x_2 < 5; 
    \]
\end{flushleft}
\vspace{-\baselineskip}
\begin{flushleft}
    \[
       0.001 < x_3 < 5; \quad 0.001 < x_4 < 5;
    \]
\end{flushleft}
\vspace{-\baselineskip}
\begin{flushleft}
    \[
        0.001 < x_5 < 5; \quad 0.001 < x_6 < 5;
    \]
\end{flushleft}
\vspace{-\baselineskip}
\begin{flushleft}
    \[
       0.001 < x_7 < 5; \quad 0.001 < x_8 < 5; 
    \]
\end{flushleft}
\vspace{-\baselineskip}
\begin{flushleft}
    \[
        0.001 < x_9 < 5; \quad 0.001 < x_{10} < 5; 
    \]
\end{flushleft}
\vspace{-\baselineskip}
\begin{flushleft}
    \[
       0.001 < x_{11} < 5; \quad 0.001 < x_{12} < 5;
    \]
\end{flushleft}
\vspace{-\baselineskip}
\begin{flushleft}
    \[
         0.001 < x_{13} < 5; \quad 0.001 < x_{14} < 5;
    \]
\end{flushleft}

\begin{figure}[htbp]
    \centering
    \includegraphics[width=0.8\textwidth]{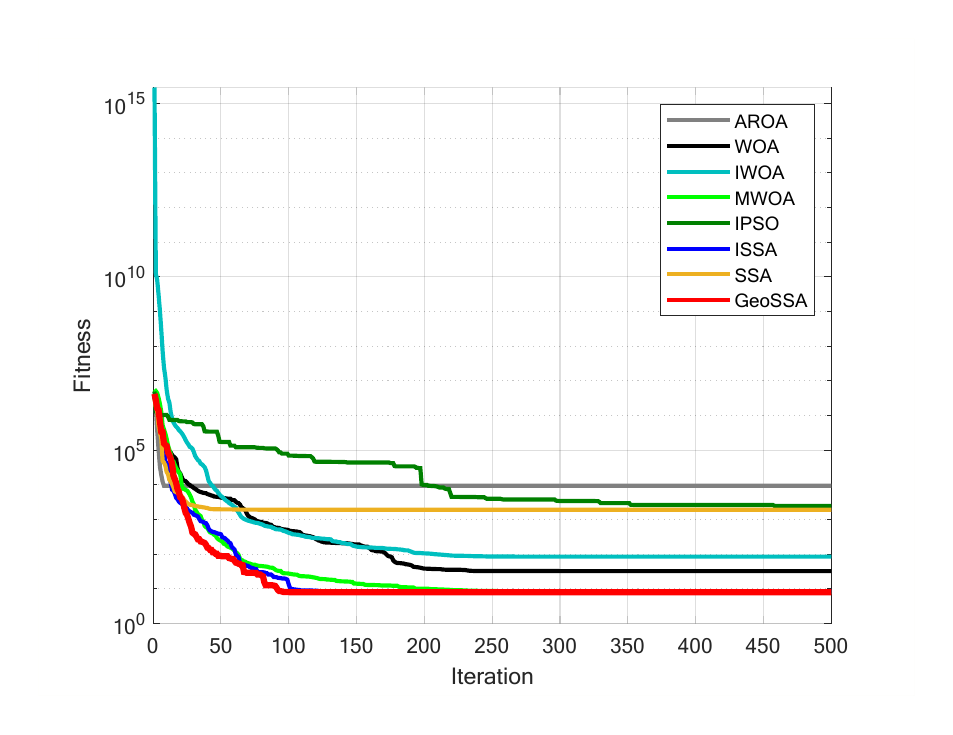}
    \caption{Iteration curves of the algorithms in solving the Industrial Refrigeration System design problem.} 
    \label{engineering42}
\end{figure}

The experimental results are presented in Fig.~\ref{engineering42} and Table~\ref{engineeringmetrics}. As shown in Table~\ref{engineeringmetrics}, in the Industrial Refrigeration System design problem, GeoSSA exhibits significantly higher stability than all other algorithms, and its optimization accuracy is the best among the compared methods. This demonstrates that GeoSSA possesses substantial advantages when addressing this type of engineering optimization problem.

\begin{sidewaystable}[htbp]
\centering
\caption{Parametric results of the algorithms in solving engineering design optimization problems.}
\begin{tabular}{ccccccccccc}
\toprule
\textbf{Problem} & \textbf{Metrics} & \textbf{AROA} & \textbf{WOA} & \textbf{IWOA} & \textbf{MWOA} & \textbf{IPSO} & \textbf{ISSA} & \textbf{SSA} & \textbf{GeoSSA} \\
\midrule
CB & Ave & 7.295368 & 7.357594 & 96.141137 & 6.894230 & 7.472589 & \textbf{6.837589} & 6.865600 & \textbf{6.837589} \\
   & Std & 0.397969 & 0.594874 & 619.597995 & 0.121083 & 0.477882 & \textbf{0.000000} & 0.198068 & \textbf{0.000000} \\
\midrule
PL & Ave & 310.340350 & 47.978511 & 117.195106 & 6.604381 & 432.790325 & 45.434656 & 50.981842 & \textbf{1.057175} \\
   & Std & 270.966089 & 91.674448 & 118.894881 & 30.383292 & 734.372867 & 74.849797 & 77.564892 & \textbf{0.000000} \\
\midrule
RN & Ave & 10.449519 & 2.719602 & 2.157282 & 0.455302 & 4.659719 & 0.657043 & 0.439659 & \textbf{0.251777} \\
   & Std & 14.145690 & 2.622095 & 1.645170 & 0.697369 & 2.191879 & 1.240103 & 0.906919 & \textbf{0.103145} \\
\midrule
IRS & Ave & 9344.680233 & 32.306197 & 84.404992 & 8.106376 & 2458.347947 & 8.302475 & 1912.056286 & \textbf{7.941763} \\
   & Std & 6050.531715 & 23.789994 & 73.284109 & 0.548582 & 5004.605495 & 0.827212 & 6019.376919 & \textbf{0.438531} \\
\bottomrule
\end{tabular}
\label{engineeringmetrics}
\end{sidewaystable}

GeoSSA achieves theoretical optimal or near-optimal results in all engineering design optimization problems. The algorithm yields the lowest average fitness values ($Ave$) and the smallest standard deviations ($Std$), fully reflecting its significant advantages in terms of accuracy, stability, and convergence speed. Therefore, the experimental results demonstrate that GeoSSA not only maintains strong global search capabilities on standard benchmark functions but also exhibits outstanding stability, precision, and versatility in engineering design optimization problems. It effectively addresses practical engineering optimization tasks characterized by complex constraints and nonlinear features.

\section{Discussion}
Future research will focus on several potential extensions of GeoSSA. These include expanding GeoSSA to multi-objective optimization through the incorporation of Pareto-front maintenance strategies; adapting it to dynamic or online optimization where objectives or constraints vary over time; and developing discrete or combinatorial variants tailored to specific encoding schemes and neighborhood structures. Moreover, integrating GeoSSA with deep learning \cite{xgb} or reinforcement learning techniques may further enhance its capability in complex policy optimization and hyperparameter tuning. In addition, leveraging parallel computing and GPU acceleration is expected to improve its scalability for large-scale optimization problems. 

\section{Conclusions}
In this study, we addressed several inherent limitations of the original Sparrow Search Algorithm (SSA) when dealing with complex optimization tasks and proposed a Geometric Sparrow Search Algorithm (GeoSSA). GeoSSA significantly enhances the overall optimization performance through three complementary improvement strategies. First, the Good Nodes Set initialization ensures well-distributed population coverage across the search domain, alleviating clustering and blind-spot issues caused by pseudo-random initialization, thereby strengthening early global exploration. Second, the Sine-Cosine Enhanced Producer Position Update strategy employs hybrid sine-cosine update dynamics together with an adaptive inertia weight to regulate the exploration-exploitation trade-off. This enables large-scale exploratory jumps in the early stages and promotes stable convergence in later iterations. Third, the Triangular Walk Enhanced Edge Sparrow Position Update introduces a scale-controlled triangular-walk perturbation for edge individuals, enhancing local search accuracy and improving the algorithm's ability to escape local optima. These three mechanisms—spanning initialization, producer update, and edge-sparrow update—each address different aspects of SSA's deficiencies while jointly improving population diversity, convergence speed, and solution quality.\par
Extensive experiments were conducted to validate the effectiveness of GeoSSA. Ablation studies on 23 classical benchmark functions demonstrated that removing any single module leads to clear performance degradation on several functions, confirming the necessity and contribution of each component. In comprehensive benchmark testing across 23 functions, GeoSSA outperformed all comparison algorithms—including AROA, WOA, IWOA, MWOA, IPSO, ISSA, and the original SSA—in terms of average fitness and standard deviation, and achieved statistically significant superiority in the Wilcoxon non-parametric test and Friedman ranking test, with an Overall Effectiveness (OE) of 95.65\%. In a UAV 3D path planning task, GeoSSA achieved the shortest and smoothest trajectories with the highest obstacle-avoidance safety and consistently demonstrated the best stability (lowest $Std$) across 30 independent runs. Furthermore, in four engineering design problems (Corrugated Bulkhead, Piston Lever, Reactor Network, and Industrial Refrigeration System), GeoSSA achieved the highest accuracy and stability, showing strong adaptability to engineering constraints and multivariable coupling, and thereby confirming its practical applicability.\par
Overall, GeoSSA provides an efficient and extensible framework for tackling complex optimization challenges in engineering applications and path planning tasks.

\section{Acknowledgment}
The supports provided by Macao Polytechnic University (MPU Grant no: RP/FCA-03/2022; RP/FCA-06/2022) and Macao Science and Technology Development Fund (FDCT Grant no: 0044/2023/ITP2) enabled us to conduct data collection, analysis, and interpretation, as well as cover expenses related to research materials and participant recruitment. MPU and FDCT investment in our work have significantly contributed to the quality and impact of our research findings.

\appendix

\section{Abbreviations}\label{secA1}

The following abbreviations are used in this manuscript:\\

\noindent 
\begin{tabular}{@{}ll}
Ave & Average fitness\\
Std & Standard deviation\\
OE &  overall effectiveness\\
CB & Corrugated Bulkhead\\
PL & Piston Lever\\
RN & Reactor Network\\
IRS & Industrial Refrigeration System
\end{tabular}

\section{Benchmark Functions}
The details of the benchmark functions used in this research are shown in Table~\ref{funcdetails}.

\begin{table}[htbp]
    \centering
    \caption{Standard Benchmark Functions~\cite{CEC}}
    \begin{tabular}{l c c c c}
    \hline
    Function & Function's Name &  Best Value \\
    \hline
    $F_1\left(x\right)=\sum_{i=1}^nx_i^2$ & Sphere &  0 \\
    $\begin{aligned}&F_{2}\left(x\right)=\sum_{i=1}^{n}\left|x_{i}\right|+\prod_{i=1}^{n}|x_{i}|\end{aligned}$ & Schwefel's Problem 2.22 & 0 \\
    $F_3\left(x\right)=\sum_{i=1}^n\left(\sum_{j-1}^ix_j\right)^2$ & Schwefel's Problem 1.2 & 0 \\
    $\begin{aligned}
        F_{4}(x) &= \max\limits_{1 \leq i \leq n} \left\{ \left| x_i \right| \right\}
        \end{aligned}$ & Schwefel's Problem 2.21 &  0 \\
    $F_{5}(x)=\sum_{i=1}^{n-1}[100(x_{i+1}-x_{i}^{2})^{2}+(x_{i}-1)^{2}]$ & Generalized Rosenbrock's Function &  0 \\
    $F_{6}(x)=\sum_{i=1}^{n}(\lfloor x_{i}+0.5\rfloor)^{2}$ & Step Function &  0 \\
    $F_{7}(x)=\sum_{i=1}^{n}ix_{i}^{4}+random[0,1)$ & Quartic Function &  0 \\
    $F_{8}(x)=\sum_{i=1}^{n}-x_{i}\sin(\sqrt{|x_{i}|})$ & Generalized Schwefel's Function &  -12569.5 \\
    $F_{9}(x)=\sum_{i=1}^{n}[x_{i}^{2}-10\cos(2\pi x_{i})+10)]$ & Generalized Rastrigin's Function &  0 \\
    $\begin{aligned}F_{10}\left(x\right)&=-20\exp\left(-0.2\sqrt{\frac{1}{n}\sum_{i=1}^{n}x_{i}^{2}}\right)-\exp\left(\frac{1}{n}\sum_{i=1}^{n}\cos2\pi x_{i}\right)\\&+20+e\end{aligned}$ & Ackley's Function &  0 \\
    $F_{11}\left(x\right)=\frac{1}{4000}\sum_{i=1}^{n}x_{i}^{2}-\prod_{i=1}^{n}\cos\left(\frac{x_{i}}{\sqrt{i}}\right)+1$ & Generalized Griewank's Function &  0 \\
    $\begin{aligned}F_{1 2}\left(x\right)&=\frac{\pi}{n}\left\{10 \sin^{2}(\pi y_{i})+\sum_{i=1}^{n-1}(y_{i}-1)^{2}[1+10 \sin^{2}(\pi y_{i+1})]\right.\\&+(y_{n}-1)^{2}\Big\}+\sum_{i=1}^{n}u(x_{i},10,100,4),\\&y_i=1+\frac{1}{4}(x_i+1)\\&u\left(x_{i},a,k,m\right)=\left\{\begin{array}{ll}k(x_{i}-a)^{m},&x_{i}>a,\\0,&-a\leq x_{i}\leq a,\\k(-x_{i}-a)^{m},&x_{i}<-a.\end{array}\right.\end{aligned}$ & Generalized Penalized Function 1 &  0 \\
    $\begin{gathered}F_{13}\left(x\right)=0.1\left\{\sin^{2}(3\pi x_{1})+\sum_{i=1}^{n-1}(x_{i}-1)^{2}[1+sin^{2}(3\pi x_{i+1})]\right.\\+(x_{n}-1)^{2}[1+\sin^{2}(2\pi x_{n})]\big\}+\sum_{i=1}^{n}u(x_{i},5,100,4)\end{gathered}$ & Generalized Penalized Function 2 &  0 \\
    $F_{14}\left(x\right)=\left[\frac{1}{500}+\sum_{j=1}^{25}\frac{1}{j+\sum_{i=1}^{2}(x_{i}-a_{ij})^{6}}\right]^{-1}$ & Shekel's Foxholes Function &  0.998 \\
    $F_{15}\left(x\right)=\sum_{i=1}^{11}\left[a_{i}-\frac{x_{1}\left(b_{i}^{2}+b_{i}x_{2}\right)}{b_{i}^{2}+b_{i}x_{3}+x_{4}}\right]^{2}$ & Kowalik's Function &  0.0003075 \\
    $F_{16}\left(x\right)=4x_{1}^{2}-2.1x_{1}^{4}+\frac{1}{3}x_{1}^{6}+x_{1}x_{2}-4x_{2}^{2}+4x_{2}^{4}$ & Six-Hump Camel-Back Function &  -1.0316 \\
    $F_{17}\left(x\right)=\left(x_{2}-\frac{5.1}{4\pi^{2}}x_{1}^{2}+\frac{5}{\pi}x_{1}-6\right)^{2}+10\left(1-\frac{1}{8\pi}\right)\cos x_{1}+10$ & Branin Function &  0.398 \\
    $\begin{aligned}F_{18}\left(x\right)&=[1+(x_{1}+x_{2}+1)^{2}(19-14x_{1}+3x_{1}^{2}-14x_{2}\\&+6x_1x_2+3x_2^2)]\times[30+(2x_1-3x_2)^2(18-32x\\&+12x_1^2+48x_2-36x_1x_2+27x_2^2)]\end{aligned}$ & Goldstein-Price Function &  3 \\
    $F_{19}\left(x\right)=-\sum_{i=1}^{4}c_{i}\exp\left[-\sum_{j=1}^{4}a_{ij}\left(x_{j}-p_{ij}\right)^{2}\right]$ & Hartman's Function 1 &  -3.8628 \\
    $F_{20}\left(x\right)=-\sum_{i=1}^{4}c_{i}\exp\left[-\sum_{j=1}^{6}a_{ij}(x_{j}-p_{ij})^{2}\right]$ & Hartman's Function 2 &  -3.32 \\
    $F_{21}\left(x\right)=-\sum_{i=1}^{5}\left[(x-a_{i})(x-a_{i})^{T}+c_{i}\right]^{-1}$ & Shekel's Function 1 &  -10.1532 \\
    $F_{22}\left(x\right)=-\sum_{i=1}^{7}\left[(x-a_{i})(x-a_{i})^{T}+c_{i}\right]^{-1}$ & Shekel's Function 2 &  -10.4029 \\
    $F_{23}\left(x\right)=-\sum_{i=1}^{11}\left[(x-a_{i})(x-a_{i})^{T}+c_{i}\right]^{-1}$ & Shekel's Function 3 &  -10.5364 \\
    \hline
    \label{funcdetails}
    \end{tabular}
\end{table}

\clearpage

\bibliographystyle{unsrt}  
\bibliography{references}

\end{document}